Accepted in "Advances in Natural Sciences: Nanoscience and Nanotechnology" on dated: 23.10.2019 (**IOP Publishing**)

# Development of Gold Tiny Particles and Particles in Different Sizes at Varying Precursor Concentration

**Mubarak Ali[1] and I-Nan Lin[2]**

[1]Department of Physics, COMSATS University Islamabad, Park Road, Islamabad-45550, PAKISTAN,

[2]Department of Physics, Tamkang University, Tamsui Dist., New Taipei City 25137, Taiwan (R.O.C.)

E-mail: mubarak74@mail.com or mubarak74@comsats.edu.pk

**Abstract**. Coalescence (or Growth) of tiny particles to larger particles has been an exciting and practical research topic in nanotechnology. This study deals with development of gold particles at varying precursor concentration in a custom-built setup. Under the tuned ratio of bipolar pulse OFF to ON time, tiny particles of different sizes and shapes develop depending on the amount of gold precursor. When compact monolayer assembly is formed on solution surface, nano energy in packet shapes bind gold atoms in own shape. Between 0.07 mM to 0.90 mM precursor concentration, tiny particles develop in both triangular and non-triangular shapes. Tiny particles of triangular shape develop in a large number at precursor concentration 0.30 mM and 0.60 mM. Hence, nanoparticles and particles also developed in different geometrical shapes. Such tiny particles pack due to the exerted force in immersing format, where made structures of smooth elements assemble to develop nanoparticles and particles of different shapes. Tiny particles of non-geometrical shapes do not assemble at a common point and their assembling develops a nanoparticle or particle of distorted or spherical shape. This occurs mainly when precursor concentration is 0.05 mM and 1.20 mM. At 50 sccm Argon flow rate, nanoparticles and particles develop in the same shapes as in case of 100 sccm. For different precursor concentrations, different colors of solutions indicate different features of nanoparticles and particles. This study clarifies the necessary concentration of precursor to develop colloids of different sizes.

## 1. Introduction

Developing different shapes and structures of colloids while processing for a certain strategy may be (is) the origin of a widespread phenomenon. The particles of controlled shape can be the strong candidates for significant applications. This is because of a great challenge to develop anisotropic particles. To assemble tiny-sized particles precisely is a goal for developing advanced functional materials. Metallic colloids having different shapes under different processing conditions may reveal the unexplored factors of their developments. When a tiny-sized particle develops by atom-to-atom amalgamation, attained dynamics of atoms can be the sole cause. For controlling (electronic) structure of atoms (configuring into tiny-shaped particles), in terms of their less or more elongation, in the course of developing shape (feature) of their large-sized particle, a tuned amount of forced energy (travelled photons) and forcing energy electrons (or electrons carrying photons) entering to solution is found to be another cause. Electron dynamics in amalgamated atoms of a certain behaviour may regulate the morphology and structure of their bigger size. However, this (electron dynamics) cannot be the case with those atoms which bind into a tiny particle due to their certain **transition** state as they should deal with infinitesimal displacement of electrons. Here, prior to binding atoms under supplied packets of nano energy, they amalgamated through their attained dynamics. On amalgamation and at first stage, atoms should configure into monolayer (not mono layer) assembly on electronically flat solution surface. On electronically flat solution surface and at second stage, from that monolayer assembly, few atoms should crop (isolate) on binding by packet of nano energy. As a result, a tiny particle (cluster of few atoms) should develop which can be referred a tiny-shaped particle. Because, it is in the same geometry (shape) as packet of nano energy. The tiny-shaped particles should be in their mono layers (not monolayers). It is expected that under variable amount of precursor concentration and Argon gas flow rate, an overall picture related to size, shape and structure of particles may be depicted.

As a matter of fact, transition state of an atom and electron dynamics of an atom are two different scenarios. In the first case, electrons should be mainly recognized to take infinitesimal displacements under varying forced energy in their atom when remaining within the occupied states (clamped energy knots). In the latter case, electrons should be mainly recognized to leave occupied states under suitable force or energy prior to reinstate states. The latter case is mainly related to the outer rings of atoms.

Several approaches have been explored in the literature to synthesize colloidal tiny-sized particles and large-sized particles, where citrate reduction method is one of the most widely adopted procedures [1]. Development of large-sized particles on likely coalescence of tiny-sized particles has been the subject of several studies [2-12]. Metal clusters behave like simple chemical compounds and could find applications in catalysis, sensors and molecular electronics [2]. Discrete features of nanocrystals and their tendency to extend into superlattices suggest ways and means for the design and fabrication of advanced materials with controlled characteristics [3]. An ordered array of nanoparticles instead of agglomeration might present new properties different from the individual particles [4]. Coalescence of nanocrystals into extended shapes has appeared to be a realistic goal [5]. Self-assembly means to design specific structure, which cannot be achieved alternatively [6]. Potential long-term use of nanoparticle technology is to develop small electronic devices [7]. Assembling of nanoparticles may be an initial effort towards the selective positioning and patterning in a large area [8]. Organization of nanometre sized building blocks into specific structures to construct functional materials and devices is one of the current challenges [9]. On assembling nanocrystals into useful structures, ‘atoms and molecules’ will be treated as materials of tomorrow [10]. Precise control on the assemblies of nanoparticles enables the synthesis of complex shapes and provides pathways to fabricate new materials and devices [11]. Coalescing nanocrystals into long-range crystals allow one to develop materials with endless selections [12]. The existing mechanistic interpretations are insufficient to explain several observations [13]. More work is required to develop in-depth understanding of metallic colloids [14, 15].

Attempts have also been made to synthesize different anisotropic particles and distorted particles in different employed plasma solution processes [16-23], where four main strategies have remained in utilization i.e., DC plasma discharge in contact with the liquid, DC glow discharge plasma in contact with the liquid, pulse plasma discharge inside the liquid, and gas-liquid interface discharge. Quite a few reaction mechanisms proposed by different groups are to be the most probable underlying mechanisms such as plasma electrons [17], hydrogen peroxide [16, 17], aqueous electrons [19] and hydrogen radicals in liquid [20]. Gold nanoplates and nanorods have been synthesized on the surface of solution while sphere-shaped particles inside the solution [20]. Again, probing matter at a length scale comparable to the subwavelength of light can deliver phenomenal optical properties [24, 25] and different phase-controlled syntheses give an improved catalytic activity of metallic nanostructures as compared to bulk ones [26, 27]. Different tiny-sized particles of gold having different attained dynamics of their amalgamation were discussed [28]. On splitting of inert gas atoms under the excessive field (population of photons having characteristics of current per unit area or volume), photons having characteristics of current enter to air/medium (open medium) by reducing their impact and become obvious to visualize the light glow [29]; when photons having characteristic of current continuously propagate through inter-state electron gaps of atoms of a suitable wire, they work as photonic current. Photons generated by the uninterrupted inter-state electron dynamics of a natural state silicon atom have their characteristics of current and energy which are being transported from one point to another through their involved force [30]. The amalgamated atoms bind under generated energy of the targeted atom while executing confined inter-state electron dynamics [31]. A separate study has discussed the mechanism of development of a **triangle-**shaped gold tiny particle and conversion of atoms of arrays into structures of smooth elements [32]; a difference between monolayer and mono layer is also discussed in that study. Carbon atoms of different states bind into grains or crystallites depending on the localized process conditions [33, 34]. Different tiny-sized particles and large-sized particles while processing gold precursor, silver precursor and their binary composition were discussed, where processing of gold solution resulted in anisotropic particles [35]. Morphology and structure of gold particles were controlled under the varying conditions of the process while employing the pulse-based electron-photon/ solution interface process [36]. While considering formation of tiny particles and their extended shapes, origin of physics and chemistry of materials was discussed [37]. Gold particles of unprecedented features developed while setting the optimized condition of the process [38]. The origin of different atoms belonging to gas and solid states along with transitional behaviours was discussed [39]. Depending on the nature of atoms of tiny-sized particles, their use can be disadvantageous as well [40].

Some details on concentration as a factor for gold nanoparticles synthesis have been discussed by the other group [16, 17]. In the molar concentration of gold precursor upto 1.0 mM, the size of gold nanoparticles were tuned in the range of few nanometres to several hundred nanometres [17]. The investigations of Patel et al. [17] concluded that hydrogen per oxide worked as the reducing agent, where gold ions converted into gold atoms. The underlying mechanism of growth of nanoparticles and particles presented in this study have never been reported before.

In the previous published work [36], the core idea was to control the shape of tiny particle, nanoparticle and particle through varying the bipolar pulse ON/OFF time and there was also the influence of pulse polarity. However, this study deeply inspires toward the role of precursor concentration in the development of tiny particle, nanoparticle and particle. The present study describes how to develop gold tiny-sized particles and their different shaped large-sized particles while varying the concentration of gold precursor. We briefly discussed the role of varying precursor concentration under fixed process parameters while developing particles of different shapes in custom-built pulse-based electron-photon/solution interface process.

## 2. Experimental details

Solid powder of $HAuCl_4$ was purchased from Alfa Aesar to obtain the aqueous solution of different molar concentrations. In brief, aqueous solution of one-gram $HAuCl_4.3H_2O$ and ~100 ml DI water was prepared in a glass bottle. This was followed by the preparation of several different molar concentrations by dissolving that amount (of prepared mixture of precursor) in DI water. The total quantity of the solution for each experiment was ~100 ml. Schematic diagram of the setup is shown in Figure 1.

A copper capillary (electron-photon source) with internal diameter ~3 mm (thickness: ~1.5 mm) was used to maintain the flow of Argon gas. At the bottom of tube, atoms of flowing Argon gas split by propagating photons having characteristic of current. A graphite rod (energy source) with a width of ~1 cm was immersed into the solution known as anode. The distance between copper capillary bottom and solution surface was ~5 mm and was kept constant in all experiments. Distance between graphite rod and copper capillary was set ~4 cm in each experiment.

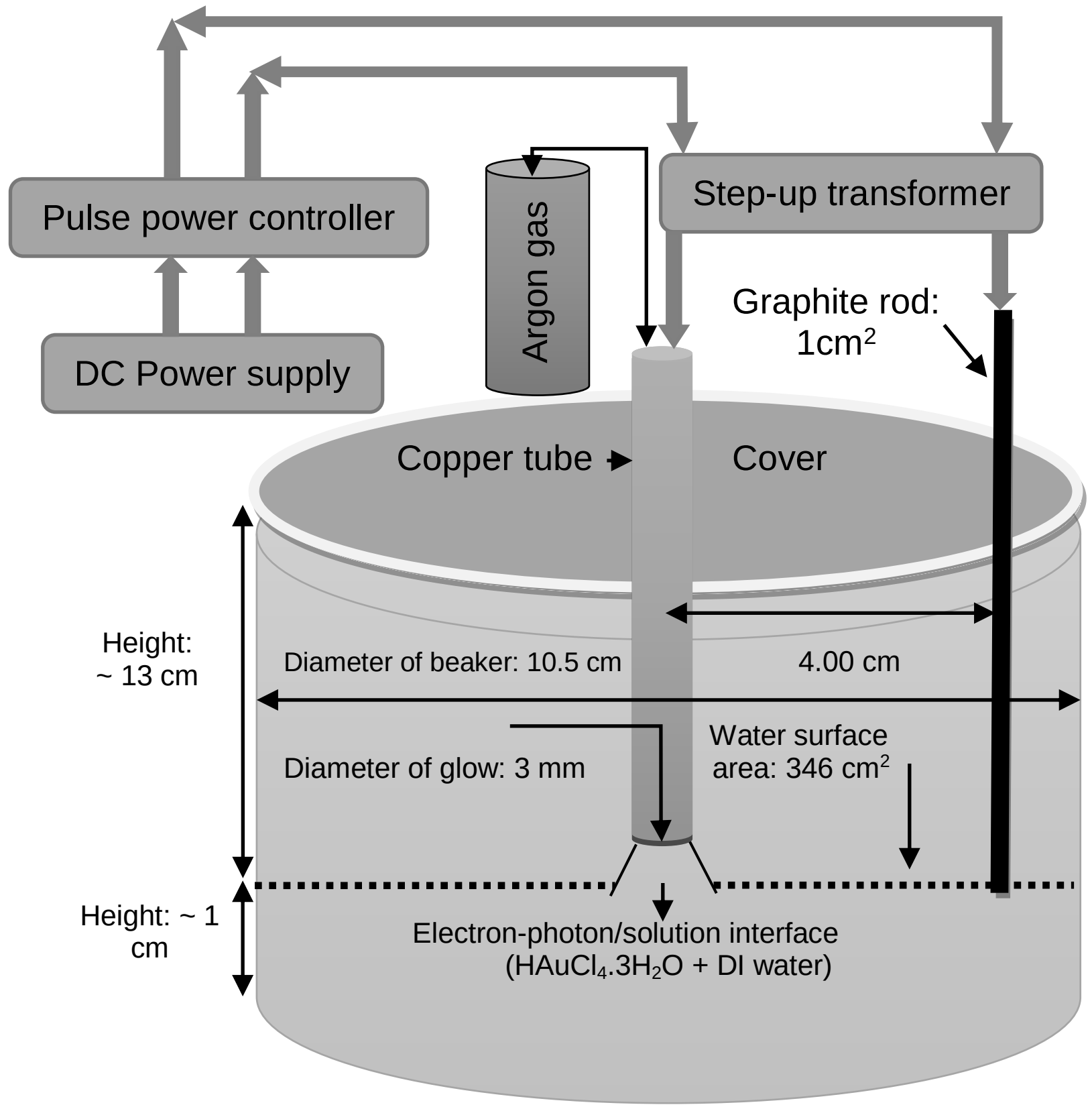

**Figure 1.** Schematic diagram of pulse-based electron-photon/solution interface process.

Bipolar pulse of fixed ON/OFF time was being controlled by the pulse DC power controller (SPIK2000A-20, MELEC GmbH; Germany). Input DC power was provided by SPIK2000A-20. Symmetric-bipolar mode of pulse power controller was employed, and equal time periods of pulses were set $t_{on}$ = 10 µsec and $t_{off}$ = 10 µsec. The input power slightly fluctuated initially. Fluctuation of input power was the maximum at the start of the process, dropped to nearly half in a second and remained almost stable in the remaining period. The splitting of inert gas atoms was automatically adjusted at the outer end forming the light glow. The light glow formed by the combined population of travelling photons and splitted electrons (and electron streams) carrying photons at the bottom of copper capillary. The stable value of the voltage was 32 volts, where current reading was noted 1.3 amp. The power was enhanced ~40 times under the application of step-up transformer. The variation of input power was noticed within 1%.

Temperature of the solution was recorded with laser-controlled temperature meter (CENTER, 350 Series). In each experiment, temperature was measured at the start (~20°C), middle (~27°C) and at the end (~37°C) of the process with ± 1°C accuracy. Different molar concentrations were prepared (~0.05 mM, ~0.10 mM, ~0.30 mM, ~0.60 mM, ~0.90 mM and ~1.20 mM), where duration of the process was set 10 minutes in each experiment. Total Argon gas flow rate was determined to be 100 sccm, which was maintained by mass flow controller passing through the copper capillary for ignition. Different molar concentrations were also prepared (~0.07 mM, ~0.10 mM, ~0.30 mM and ~0.60 mM) at 50 sccm Argon gas flow rate, where other parameters were kept the same as in case of processing solutions at 100 sccm Argon gas flow rate.

Copper grid covered by carbon film was used and samples were prepared by dip-casting. Samples were placed into Photoplate degasser (JEOL EM-DSC30) for ~24 hours to eliminate moisture. Bright field transmission microscope (BF-TM) and high-resolution transmission microscope (HR-TM) images were taken by the application of microscope known as HR-TEM (JEOL JEM2100F; operated at 200 kV). Selected area photon reflection (SAPR) patterns known as SAED patterns were also taken by HR-TM.

## 3. Results and discussion

At precursor concentration 0.05 mM, sphere-shaped and less distorted nanoparticles developed as shown by different BF-TM images (a-d) in Figure 2. Nanoparticles appear to have developed due to the coalescences of very small tiny particles. They hardly show any features of geometrical shapes, where they appeal to distorted nanoparticles mainly.

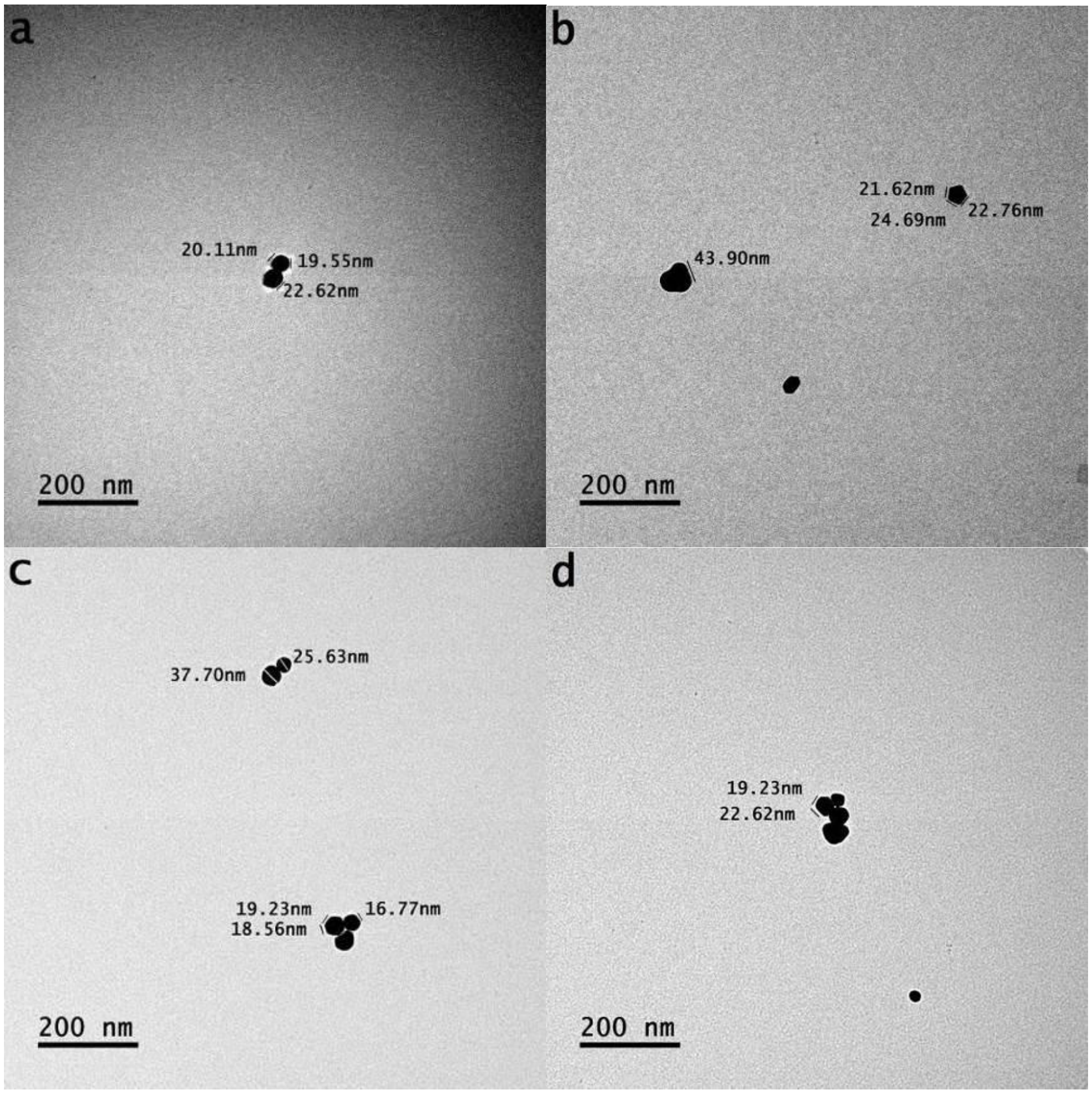

**Figure 2.** (a-d) BF-TM images of nanoparticles showing various less distorted shapes, precursor concentration 0.05 mM and Argon gas flow rate 100 sccm.

Although, at very low precursor concentration, the colloidal entities (nanoparticles) are smaller in size to tiny-sized particles developed for higher concentration of gold precursor, but they developed on the coalescences of tiny-sized particles having atoms in a few numbers. So, that is why, they are also called nanoparticles. Nanoparticles are at second place in developed colloidal entities as particles. Both nanoparticle and particle, in any shape, are not the first-hand developed colloidal entities. Generally, on the approximate scale of length when a side length is around 100 nm, it is considered a nanoparticle and when it is greater, it is considered a particle. An anisotropic particle has structure different to that of the distorted one.

In case of precursor concentration 0.05 mM, the average size of nanoparticle is between 20 to 25 nm. This indicates that tiny-sized particles developed in the formation of such nanoparticles only constituted a few gold atoms. But, on increasing the precursor concentration from 0.05 mM to 0.10 mM, the average size of particles also grew larger, where many of them developed in geometrical shapes as shown by BF-TM images in Figure 3 (a-c).

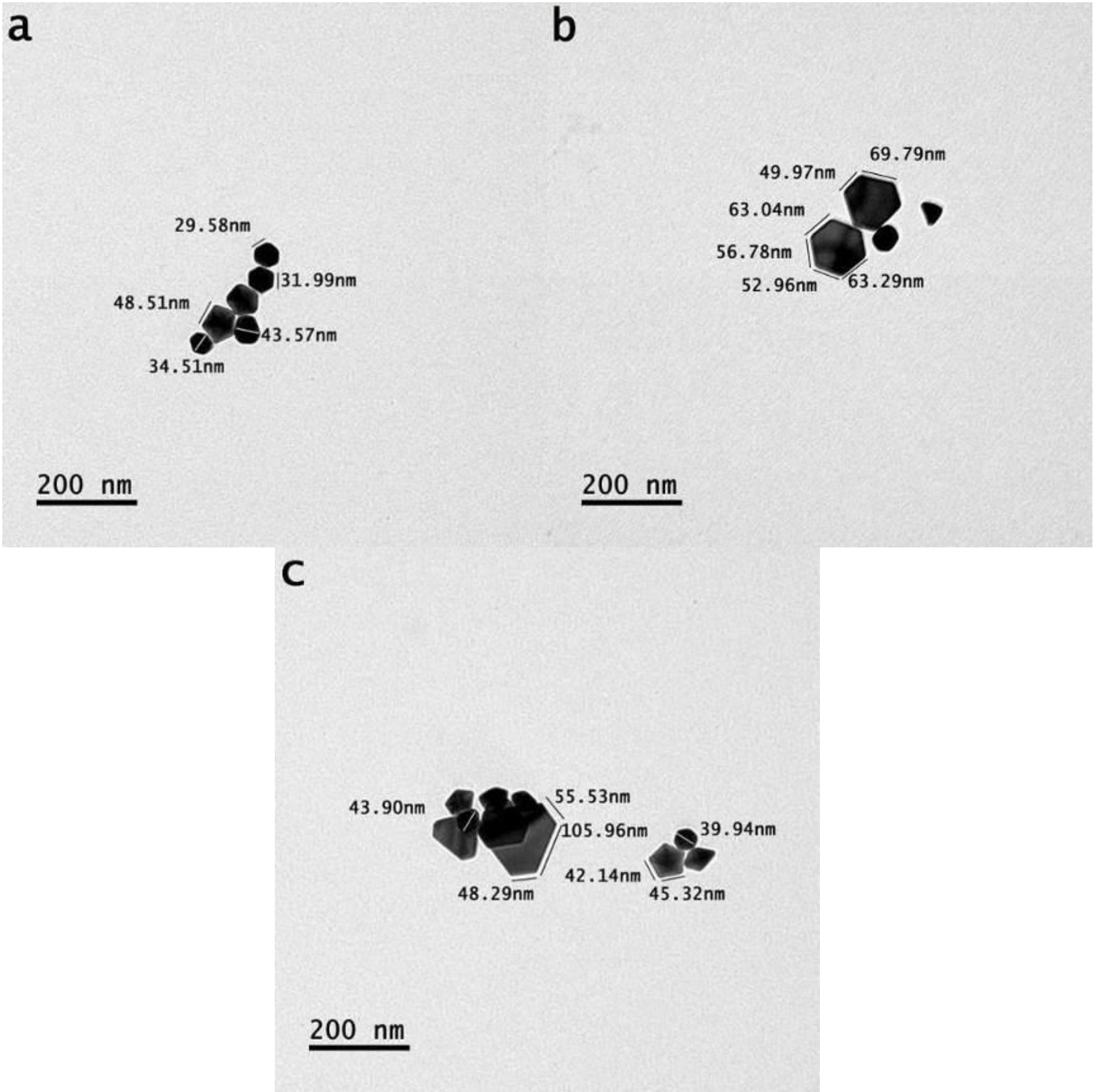

**Figure 3.** BF-TM images of nanoparticles showing both geometrical and distorted shapes; precursor concentration 0.10 mM and Argon flow rate 100 sccm.

The number of amalgamating atoms per unit area has increased. So, the size of tiny particle has also increased. Here, tiny particles also developed in triangular shape in quite a large number. Hence, their assembling under the certain mechanism also resulted in the development of **geometric**-shaped nanoparticles.

At precursor concentration 0.30 mM, the average size of particles having different geometrical shapes is increased further. The BF-TM images of triangle-, hexagon-, isosceles trapezoid-, rhombus-, pentagon-, rod- and bar-shaped particles are shown in Figure 4 (a) & (b). The particles show high aspect ratios and their increase in size is related to increase in size of tiny particles. As the size of tiny particles having triangular shape is increased further, the assembling results in the development of both nanoparticles and particles.

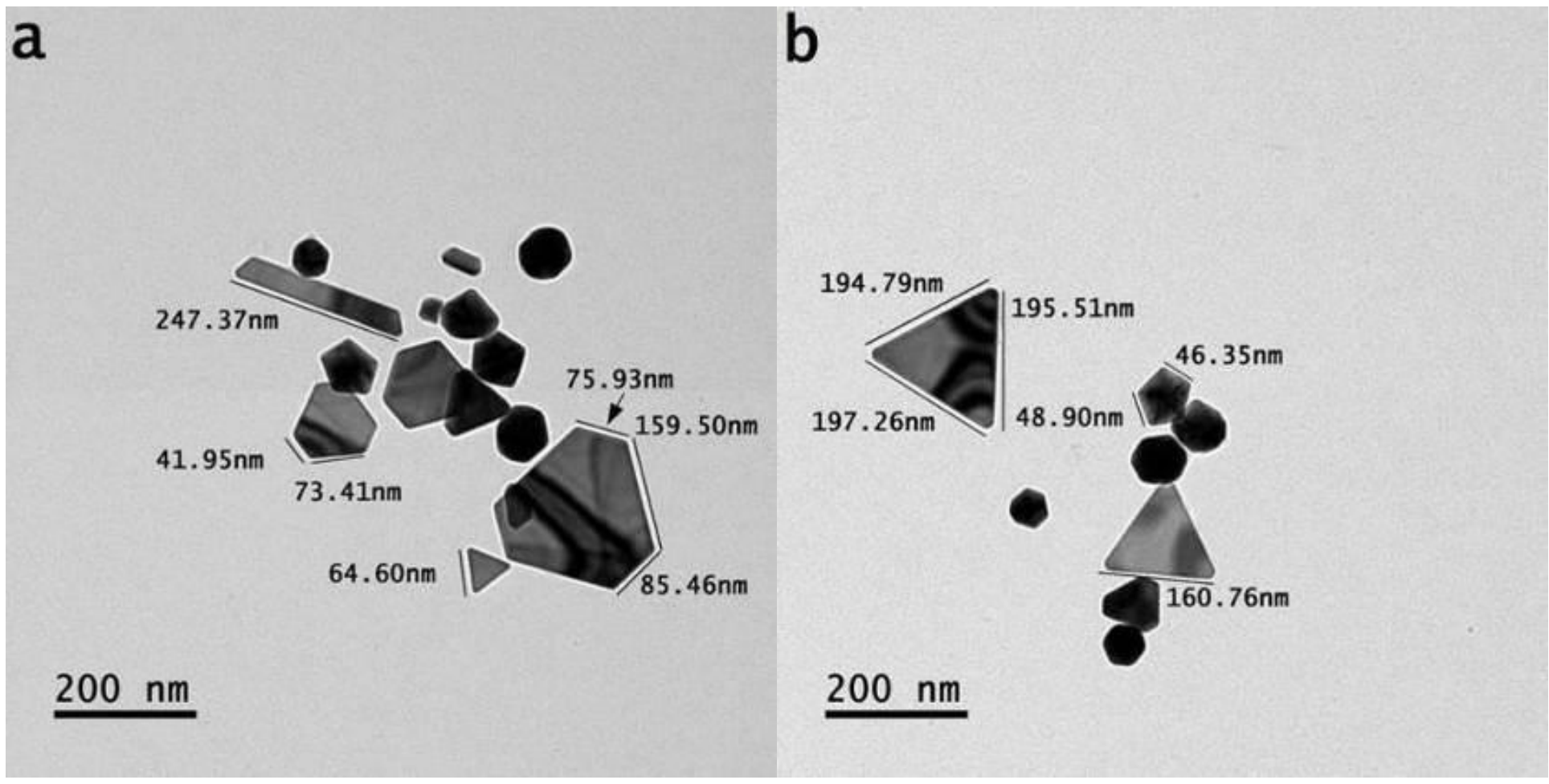

**Figure 4.** (a) & (b) BF-TM images of nanoparticles and particles developed in various geometrical shapes and distorted shapes; precursor concentration 0.30 mM and Argon gas flow rate 100 sccm.

Several high aspect ratio shapes are shown by BF-TM images in Figures 5 (a-i). Each image shows a unique geometrical shape of the particle. In Figure 5 (g), difference in the lengths of sides of particles (triangle and hexagon-shaped particles) is within the margin of an atom or few atoms. This indicates that tiny particles of same size were packed. These particles developed by the assembling (packing) of tiny-shaped particles to all sides at uniform rate. In some cases, the particles bind *via* their sides (Figure 5h), and, in other cases, they are overlaid by one another (Figure 5i), but they are related to distorted particles. SAPR patterns of particles in Figure 5 (a-f) are shown in Figure S1 (A-F). These show ordered structures of the one-dimensional (1D) shapes in case of rod like particles and multi-dimensional (MD) shapes in case of triangle/hexagon like particles.

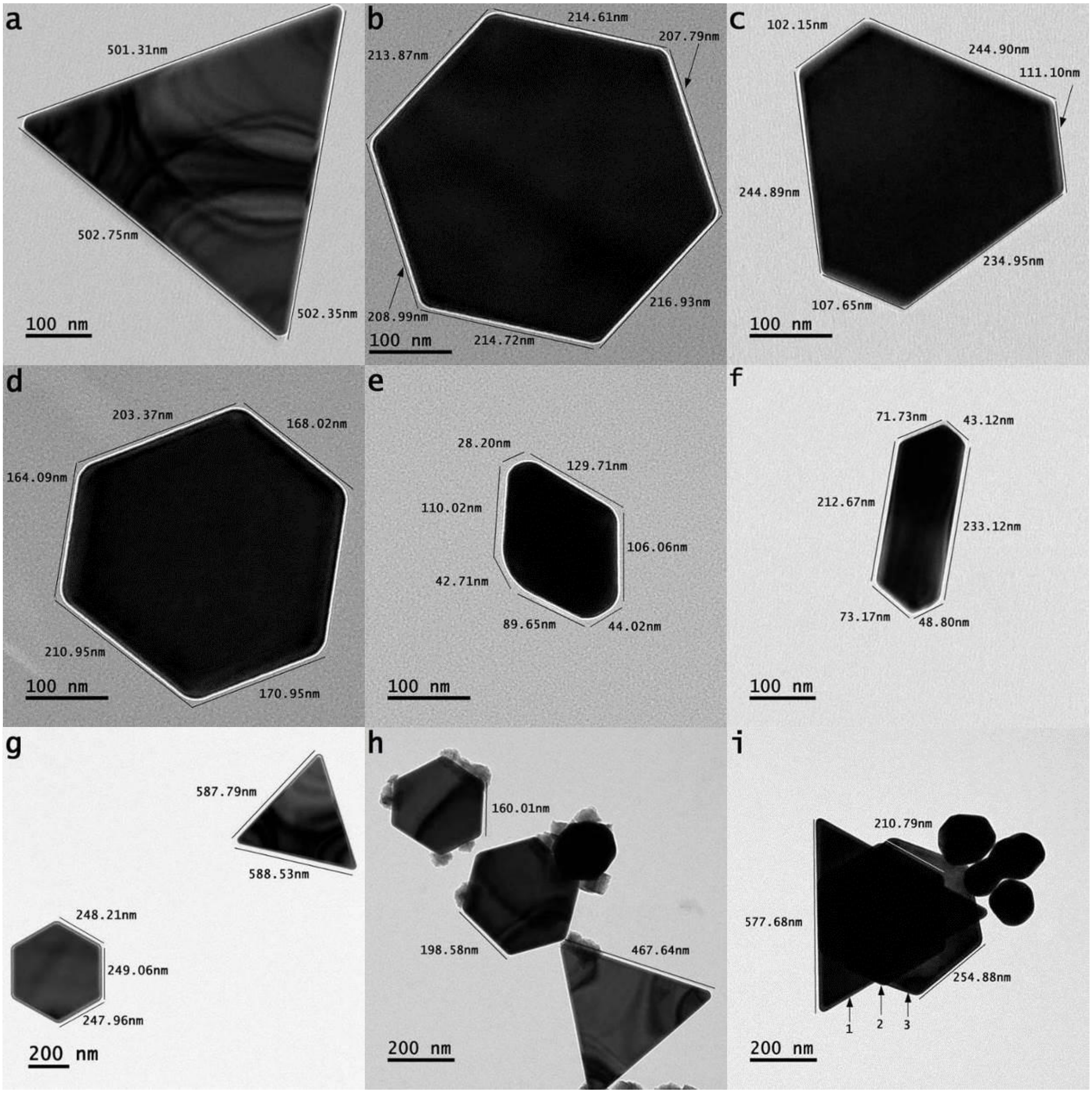

**Figure 5.** (a-i) BF-TM images of particles showing both geometrical shapes and distorted shapes; precursor concentration 0.60 mM and Argon gas flow rate 100 sccm.

For precursor concentration 0.90 mM, particles of different geometrical shapes are shown by BF-TM images in Figure 6 (a-h), which indicate the similar features as in case of particles (and nanoparticles) developed at precursor concentrations 0.10 mM, 0.30 mM and 0.60 mM. However, particles developed at 0.90 mM possess a low aspect ratio also. In Figure S2 (A-F), SAPR patterns also show ordered structures of the particles.

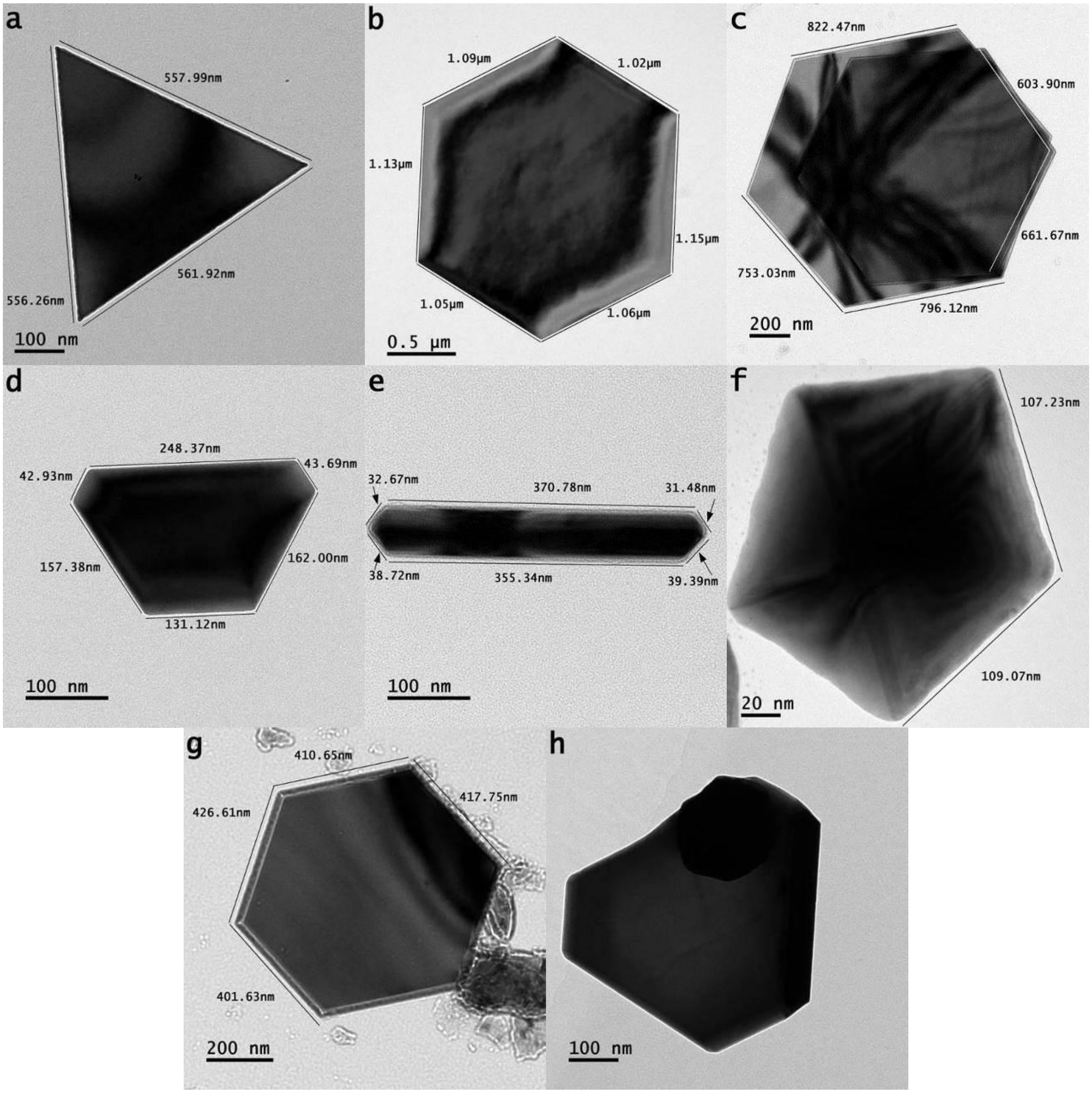

**Figure 6.** (a-h) BF-TM images of particles showing both geometrical and distorted shapes; precursor concentration 0.90 mM and Argon flow rate 100 sccm.

At 1.20 mM, very large sized tiny particles are packed under the influence of forces of **uneven** behaviors which results in the development of highly distorted particles as shown by various BF-TM images in Figure 7 (b-j). Only the particle of hexagonal shape shows anisotropy as shown in Figure 7 (a). This indicates its development at later stage of the process. However, the elongation of atoms is not appeared to be uniform in case of distorted particles shown in Figures 7 (c) and 7 (d). The very large-sized packed tiny particles do not indicate the uniform elongation of their atoms as the shape of developed particle is distorted which is shown in Figure 7 (d). A flower like shape of distorted particle is shown in Figure 7 (e) and several particles of identical features are shown in Figure 7 (i). In Figure 7 (e), an average size of tiny particle is 50 nm, which causes development of highly distorted particles. SAPR patterns of particles (shown in Figure 7a-d) are discussed in the section of supplementary materials (in Figure S3). Except for the particle shown in Figure 7 (a, b), the structures of particles are completely in disorderliness. However, the morphology of particle shown in Figure 7 (b) is rough.

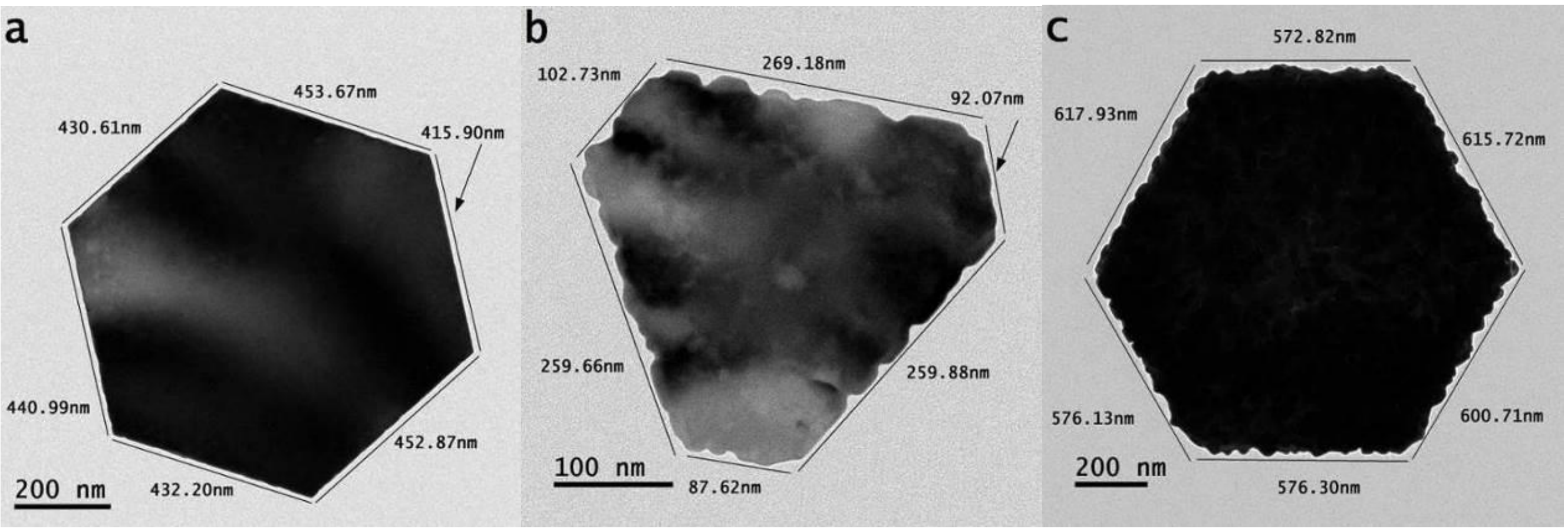

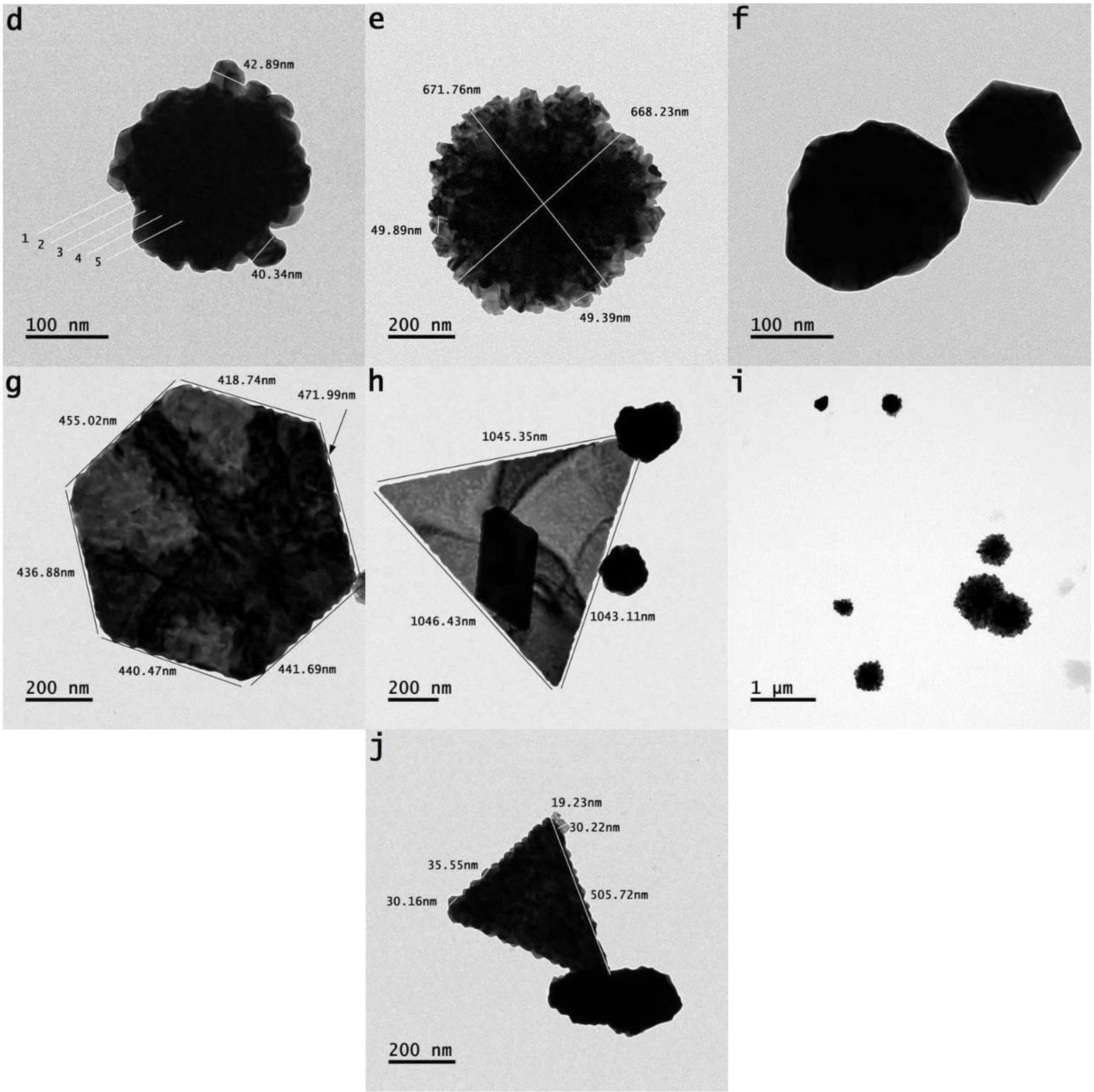

**Figure 7.** (a-j) BF-TM images of particles showing distorted particles (except in 'a, b'); precursor concentration 1.20 mM and Argon flow rate 100 sccm.

As long as atoms are bound to form first-hand cluster, molecule, nano crystal or tiny particle, it is related to a tiny-sized particle. So, a first-hand union of atoms can be in a less number or in a large number. This mainly depends on the initial used concentration of the precursor. As an evidence, in Figure 7, the signs of coalesced tiny-sized particles in developed particles are obvious, where the greater amount of initial concentration (1.20 mM) of gold precursor was used. So, the developed first-hand tiny particles are in their very large size. Because of their first-hand developing, they, therefore, are also termed as tiny-sized particles. The average size of tiny particle reached the limit of 40 nm in the particle shown in Figure 7 (d) and 50 nm in the particle shown in Figure 7 (e). However, these sizes are maximum for these tiny particles. They are related to tiny-sized particles as well because of developing under the first-hand coalescences of atoms, but in a large number. In addition to develop under the first-hand coalescences of atoms, they have the scope of being self-driven in the available force to assemble (without loss of time) to develop a nanoparticle or particle. But coalescences of such tiny-sized particles developed nanoparticle or particle having distorted features of their shapes. However, when the coalescence of such tiny-sized particles is in a less number, they developed a nanoparticle of distorted shape and when the coalescence of such tiny-sized particles is in a large number, they developed a particle of distorted shape.

The colors of processed solutions at different molarities of precursor concentration are shown in Figure 8a (left to right: 0.05 mM, 0.10 mM, 0.30 mM, 0.60 mM, 0.90 mM and 1.20 mM). Besides 100 sccm, solutions were also processed at 50 sccm Argon gas flow rate and their colors are shown in Figure 8b (left to right: 0.07 mM, 0.10 mM, 0.30 mM and 0.60 mM). A different color of each solution is related to the overall size and shape of particles along with their quantity. Appearance of distinctive color of a colloidal solution processed at different molar concentration is related to the scheme of inter-state electron gaps of elongated atoms processed by structures of smooth elements of different shapes particles and the number of their overall quantity. These require further investigations at a large scale.

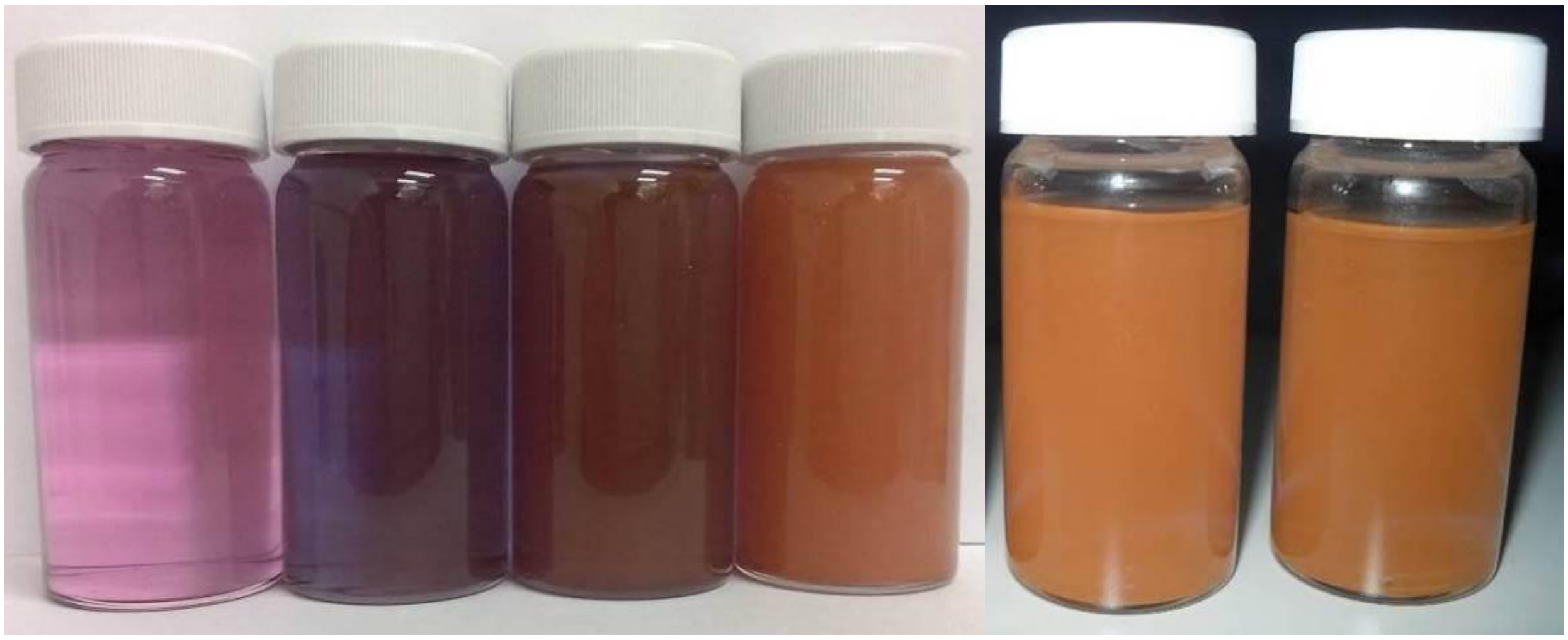

**Figure 8a.** Different color of solutions processed under different molar concentration of precursor (0.05 mM, 0.10 mM, 0.30 mM, 0.60 mM, 0.90 mM and 1.20 mM, left to right) and Argon flow rate 100 sccm.

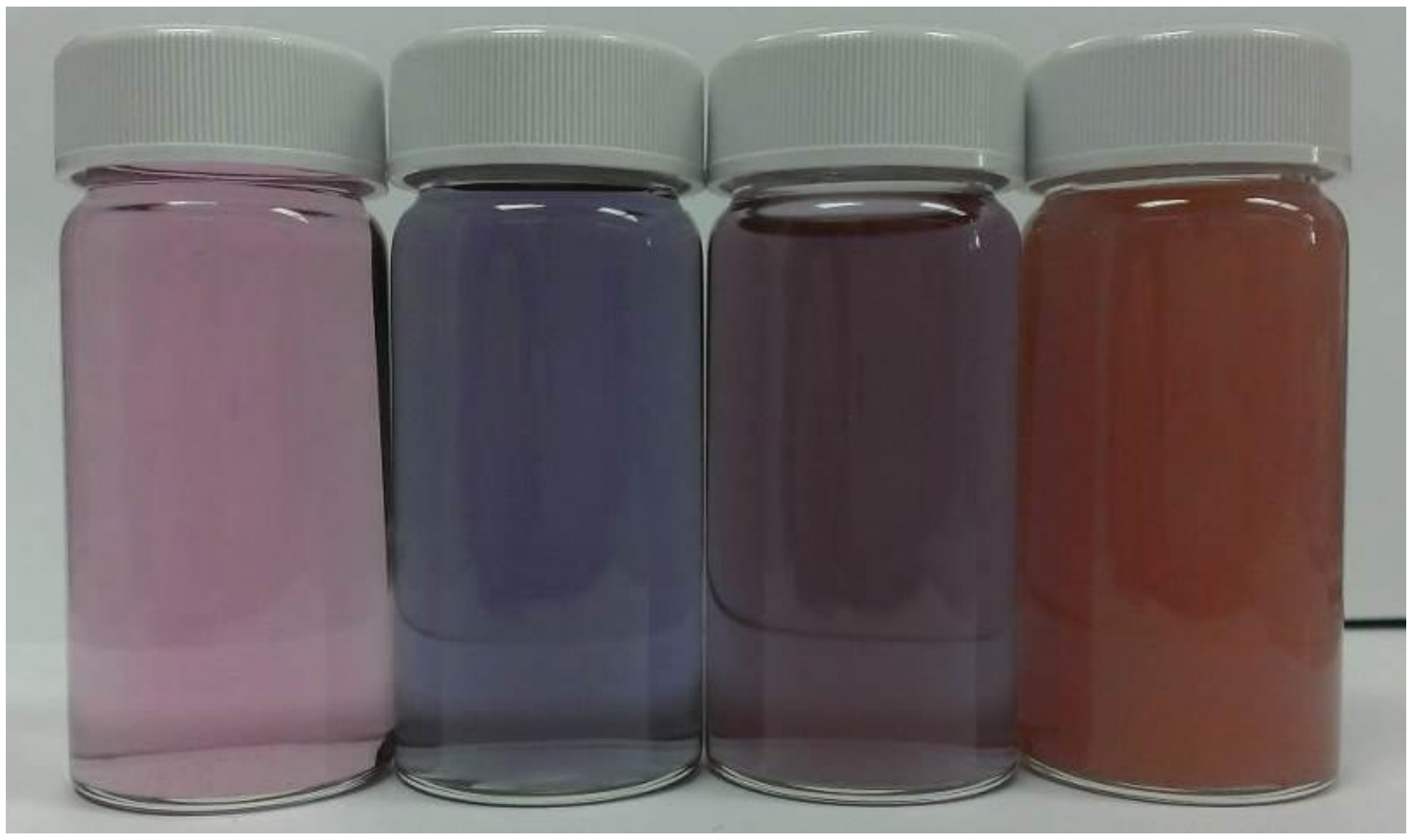

**Figure 8b.** Different color of solutions processed under different molar concentration of precursor (0.07 mM, 0.10 mM, 0.30 mM and 0.60 mM, left to right) and Argon flow rate 50 sccm.

The BF-TM images of nanoparticles and particles developed for various concentrations when Argon gas flow rate was 50 sccm are shown in Figures S4-S7. Distorted particles as well as anisotropic particles developed at 50 sccm show identical features to the ones developed at 100 sccm. The nanoparticles and particles developed at different concentrations of gold solution when processed at 50 sccm Argon gas flow rate retain their shapes as in case of nanoparticles and particles developed at 100 sccm Argon gas flow rate. Many of the nanoparticles developed in the form of geometrical shapes at precursor concentration 0.07 mM as shown in Figure S4 (a-c), in different BF-TM images.

The number of nanoparticles and particles having geometrical shapes is further increased at precursor concentration 0.10 mM. As shown in Figure S5 (a-d), there are many nanoparticles and particles which developed in geometrical shapes. Nanoparticles of very small size and very large size are found in a large number as shown in Figure S5 (a) and Figure S5 (c).

The size of nanoparticles and particles is increased further when higher concentration of gold solution is processed, where both anisotropic and distorted particles developed as shown in Figure S6 (a-c). Distorted nanoparticles and particles are assembled to an anisotropic particle to fill the vacant space through available forces as shown in Figure S6 (a). Both distorted nanoparticles and particles are assembled also to fill their vacant regions due to the **uneven** behaviors of forces as shown in Figure S6 (c).

Different BF-TM images of nanoparticles and particles developed at 50 sccm Argon gas flow rate when the precursor concentration was 0.60 mM as shown in Figure S7. Some of the nanoparticles and particles developed having precision of an atom in their sides, or a few atoms, for example, a triangle-shaped particle shown in Figure S7 (g). The different hexagon-shaped nanoparticles (or particles) shown in Figure S7 (c) also indicate the lengths of their sides in highly controlled precision.

Several different shapes of nanoparticles and particles are shown in Figure S8 (a). HR-TM image of one vertex of a triangle-shaped nanoparticle encircled in Figure S8 (a) was captured. The magnified view of that HR-TM image is shown in Figure S8 (b). In this image, width of elongated atoms of each array made in a structure of smooth element is ~0.12 nm (but not = 0.12 nm). Nearly same is the case with their inter-spacing distance (~0.12 nm (but not = 0.12 nm), which is also shown in Figure S8 (b). Both width and inter-spacing distance are measured with original scale marker.

Nanoparticles and particles that developed at Argon gas flow rate of 50 sccm show intensive contrast of dark color as compared to different featured nanoparticles and particles synthesized at Argon gas flow rate of 100 sccm. This is more relevant to elongation of atoms in different tiny-shaped particles developing nanoparticles and particles of geometrical shapes. At Argon gas flow rate of 50 sccm, elongation rate of gold atoms is slightly less (at sub-angstrom level) due to lower concentration of flowing Argon gas. Therefore, splitted electron streams of flowing inert gas atoms (under lower concentration) impinge on the underneath matter (gold atoms) at a lower rate. However, due to the entrance of less population of photons and electron streams (resulted under splitted flowing inert gas atoms) in the solution, the average size of nanoparticles and particles also becomes smaller as compared to those developed at 100 sccm Argon gas flow rate. Under different Argon gas flow rate, it is expected that inter-state electron gaps of elongated atoms of structures of smooth elements of nanoparticles and particles varies within sub-angstrom (≤ 0.10 Å). This difference is because of more or less elongations of the atoms of their structures of smooth elements. In addition to that, anisotropic nanoparticles and particles developed at 100 sccm Argon gas flow rate can have smaller circular shape of printed white spots of intensity in their SAPR patterns as compared to anisotropic nanoparticles and particles developed at 50 sccm Argon gas flow rate. Generally, nanoparticles and particles developed at 50 sccm Argon gas flow rate have dark black texture of their surfaces in BF-TM images, whereas those developed at 100 sccm show only black texture of their surfaces in BF-TM images.

It is observable in the BF-TM images of nanoparticles and particles developed at 50 sccm and 100 sccm Argon gas flow rate that they reveal not only a bit difference in size, but also in the solutions with different colors. So, by varying the inter-state electron gaps of elongated gold atoms (or by atoms of other suitable elements), it is possible to report the underlying science of different colors. However, many studies are required to investigate the influence of different Argon flow rates on the structures and, ultimately, on their shapes and sizes also. This will possibly depict the overall picture of splitting light into different colors through different colloidal matters processed under different conditions.

Packing trends of tiny-shaped particles can be depicted and drawn from their differently developed anisotropic particles. Assembling angles can be extracted from the distributed intensity spots in SAPR patterns of their particles.

On amalgamation of atoms at set precursor concentration, they developed in different tiny-sized particles under the supply of packets of nano energy. At the lowest concentration of precursor (0.05 mM), very few gold atoms were available at the solution surface. However, the packet of each nano energy possesses the same size and shape under set tuned ratio of pulse OFF to ON time as given in the section of experimental details. The atoms underneath packets of nano energy are insufficient to form their reasonable sizes. Their average size remained ~1.3 nm with no specific geometry. Atoms of assembly do not form compact monolayer at air/solution interface. Thus, at 0.05 mM (and when ratio of pulse OFF to ON time =1), atoms do not develop tiny particles in a triangular shape (Figure 9$a_1$) because atoms do not align in arrays of their tiny clusters. Thus, they do not elongate for developing structures of smooth elements as shown in Figure 9 ($a_2$). They are packed under the exertion of force in **uneven behavior**. This results in the development of less distorted or sphere-shaped nanoparticle as shown in Figure 9 ($a_3$).

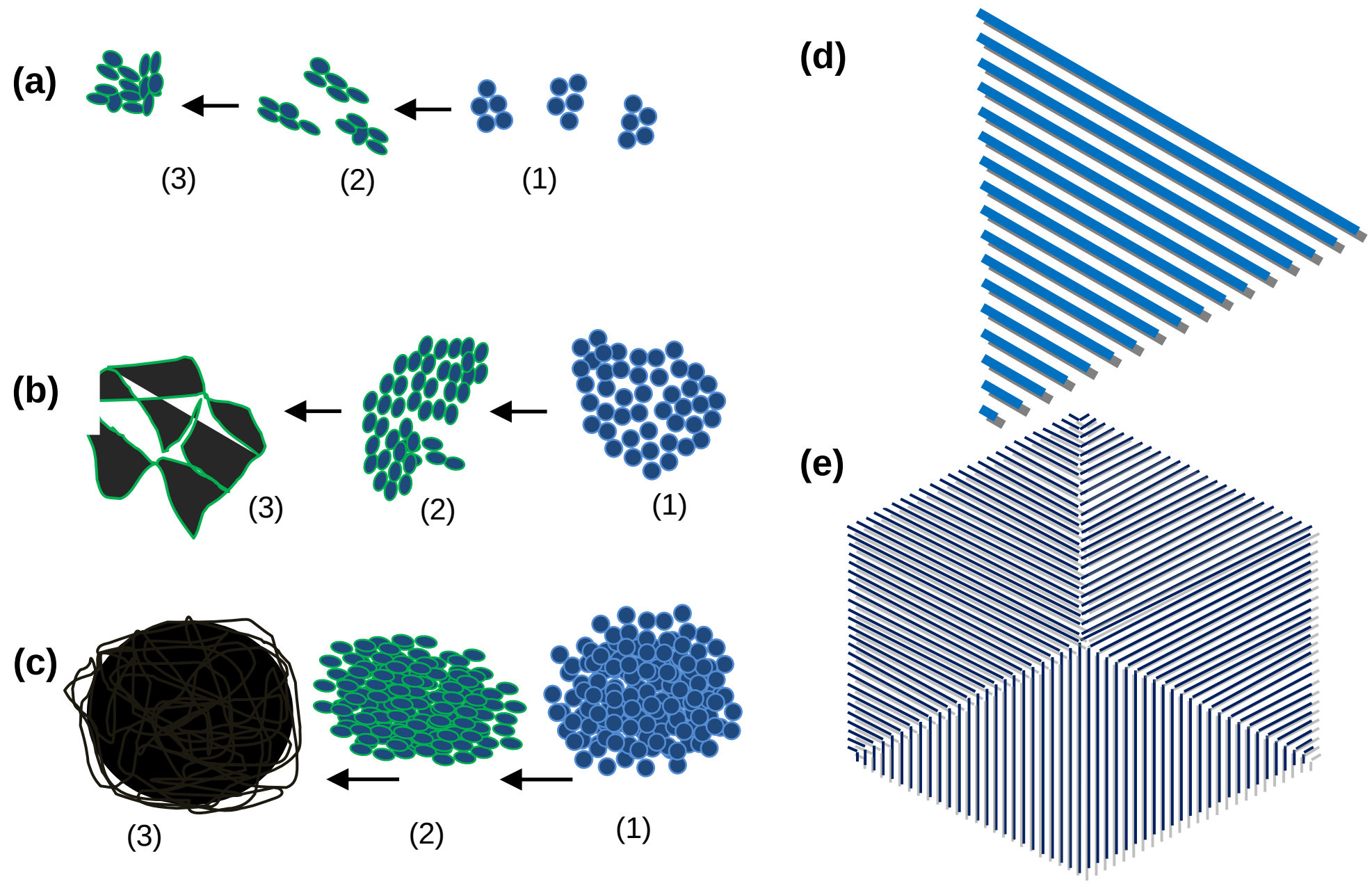

**Figure 9.** ($a_1$) less disordered tiny particles, ($a_2$) deformation of atoms of 'less disordered tiny particles', ($a_3$) less distorted particle, ($b_1$) moderately disordered tiny particle, ($b_2$) deformation of atoms of 'moderately disordered tiny particle', ($b_3$) partially distorted particle, ($c_1$) highly disordered tiny particle, ($c_2$) deformation of atoms of 'highly disordered tiny particle', ($c_3$) highly distorted particle, (d) development of particle in a triangular shape having the MD shape in the region of three dimensions and (e) development of particle in a hexagonal shape having the MD shape in the region of six dimensions.

At fixed bipolar pulse ON/OFF time while increasing the precursor concentration from ~0.07 mM to ~0.90 mM, many tiny particles developed in a triangular shape. Their size increased on increasing precursor concentration. However, for intermediate precursor concentrations (~0.07 nm to ~0.90 mM), tiny-sized particles also developed in a non-triangular shape in certain regions of the solution. They are termed as moderately disordered tiny particles as shown in Figure 9 ($b_1$). Such tiny particles do not deal with packing under exertion of force in an even way. Thus, not all the atoms of such tiny particle elongate uniformly. They are termed as moderately distorted tiny particles as sketched in Figure 9 ($b_2$). So, such tiny particles packed under exertion of force in an uneven way. Their packing developed partially distorted nanoparticles and particles as shown in Figure 9 ($b_3$).

At 1.20 mM, precursor concentration is very large, and assembly developed on solution surface does not deal with the ordering of gold atoms from the start of the process. Due to the higher amount of gold atoms on solution surface, it resulted in the development of an assembly of disorderliness instead of perfect monolayer. However, the packets of nano energy owned their same size and shape as in case of processing solutions of lower concentration. Therefore, at initial stage of the process, all the developed tiny-sized particles are highly disordered as they are quite large. The highly disordered tiny particles do not develop in a triangular shape. Hence, their structures do not undertake arrays of atoms. A highly disordered tiny particle is shown in Figure 9 ($c_1$), where groups of atoms (total atoms: 171) configured along different sides. Thus, atoms of highly disordered tiny particle elongated and deformed along the different orientations, where forces exerted along the poles of electrons in a highly uneven manner. Deformation of atoms of ‘highly disordered tiny particle’ in different orientations is shown in Figure 9 ($c_2$). Such large-sized tiny particles do not pack under the even exertion of force to electrons. The packing and assembling of such highly distorted tiny particles developed a highly distorted particle as shown in Figure 9 ($c_3$).

A ‘tiny particle’ is referred to a ‘tiny-sized particle’. A tiny particle of both small and large size is a tiny-sized particle. However, in some sentences, it does not appeal to its extracted meaning. So, to avoid any ambiguity, a ‘tiny-sized particle’ is used frequently in its place. Nevertheless, when a tiny particle has very large size, it is a tiny particle of very large size also and when a tiny particle has very small size, it is a tiny particle of very small size also.

Particles of different geometrical shapes developed one at a time. They developed at the centre of light glow (common point) on solution surface in the course to maintain their development procedure. The shape of particle depends on the number of packing of tiny-shaped particles at one time arriving from certain regions (zones) of the solution surface. Both the particles shown in Figure 9 (d) and (e) developed at the centre of light glow. In both cases, packing of tiny-shaped particles remained uninterrupted to develop structures of smooth elements in the hexagon and triangle-shaped particles. In the triangle-shaped particle, the orientation of structures of smooth elements for each face is made at ~120°. As shown in Figure 9 (d), each face of the particle having MD shape has uniform placement of structures of smooth elements having their equal widths and inter-spacing distances. This is the case in each composed mono layer of that particle. The orientation of a structure of smooth element for each face is along the ~120° angle, which is regular throughout. This particle forms three dimensions. In the hexagon-shaped particle, the orientation of structures of smooth elements is framed for six different faces. As shown in Figure 9 (e), each face of the particle of MD shape formed its uniform placement of structures of smooth elements by means of required forced exertions mainly in the surface format and partially in the space and grounded formats. The orientation of a structure of smooth element for each face is along the ~60° angle, which is regular throughout. That particle belonged to MD shape having its six dimensions. Thus, the flat region of each face measures approximately one radian. Here, the width of each structure of smooth element and inter-spacing distance also remained approximately the same. Thickness of each structure of smooth element appeared to be the same as that of the resultant width of each elongated gold atoms of array. Before assembling tiny-shaped particles to nucleate and develop these particles of MD shapes or 1D shapes, each tiny particle developed in the form of block of joined triangle-shaped tiny particles under the bipolar pulse, where they separated under the exertion of force. A detailed study on the separation of joined tiny-shaped particles is given elsewhere [33], where a structure of smooth element is formed on elongating and binding atoms of array. A single elongated atom also represents a structure of smooth element. The shapes of different anisotropic particles deal with their structures having smooth elements, where they get formed by the more or less elongation of atoms [41]; in MD shapes, atoms elongate more to form smooth elements and in 1D shapes, atoms elongate less to form smooth elements. So, anisotropic nanoparticles and particles deal with structures of smooth elements instead of structures of atoms. Therefore, mono layers of structures of smooth elements form different orientated faces to develop nanoparticle or particle of 1D shape or MD shape. This is through the precise assembling of structures of smooth elements belonging to different tiny particles of triangular shapes.

A structure of smooth element develops when transitional **behavior** atoms in array remain bound while elongating at the same rate from both sides of centers under exertion of force to their electrons along the opposite poles [31]. So, the structures of smooth elements are mainly related to the tiny particles having their triangular shapes. Elongation of single gold atom is discussed under the equal rate along the both sides (poles) from its centre [28]. In the elongation of an atom, electrons of both sides from their (atom) centre orientate from the lateral orientation to the adjacent orientation [36].

Under very high concentration of gold precursor (1.20 mM), average size of tiny particles was 50 nm at the start of the process and on prolonging the process time, tiny particles resulted in a decreased size as discussed elsewhere [28]. Therefore, the anisotropic particle shown in Figure 7 (a) is due to the tiny particles of smaller size developed at the later stage of the process. This indicates that, by increasing the process duration, the favorable conditions prevail. The tiny non-triangular **shape** particles changed into triangular shape under the favorable conditions of the process. Therefore, initial concentration of precursor is not the only parameter controlling the size and shape of tiny particles. Size and shape of tiny particles as well as of nanoparticles and particles depend on time to time change in the precursor concentration, too.

Simultaneous assembling of structures of smooth elements of two triangle-shaped tiny particles at the centre of light glow from opposite sides is almost along the same axis, so they developed a mono layer of developing particles of 1D shapes. A pentagon-shaped particle and a hexagon-shaped particle are related to five dimensions and six dimensions respectively. However, such dimensional shapes are not possible when gold atoms result in the emergence of structure in the original format by executing confined inter-state electron dynamics [31]. So, developing such types of large-shaped particles showing anisotropy, which are also the part of extensive debate in the literature, is under the development of tiny-shaped particles.

The approximate percentage of anisotropic and distorted shapes developed at different molar concentrations of gold is represented in a chart form as shown in Figure 10. As shown in Figure 10, at intermediate range of precursor concentration, many particles developed in geometrical shapes. They are approximately more than 50% in number as shown in Figure 10, which is not the case at very low and very high concentrations of solution, where mainly distorted nanoparticles and particles developed in a large number. However, for the precursor concentration 0.10 mM, approx. 25% nanoparticles and for the precursor concentration 0.90 mM, approx. 35% particles developed in geometrical shapes (in Figure 10). The average size of both nanoparticles and particles having different features can be estimated from the side lengths of nanoparticles and particles shown in their respective figures. For each processed concentration of gold precursor, side lengths of the resultant nanoparticles and particles are labelled in their respective images.

In the relevant Figures shown here, data from some BF-TM images might not represent the whole nanoparticles and particles. Therefore, distribution of the whole nanoparticles and particles for each concentration should be further investigated. In Figure 10, the presented distribution of nanoparticles and particles (of anisotropic and distorted shapes) is based on the analysis of BF-TM images given in this work and allied works of the authors [36-38]. The distribution does not infer the results of absorption or Zeta potential analysis. It is possible to perform different statistical analyses of nanoparticles and particles including standard deviation.

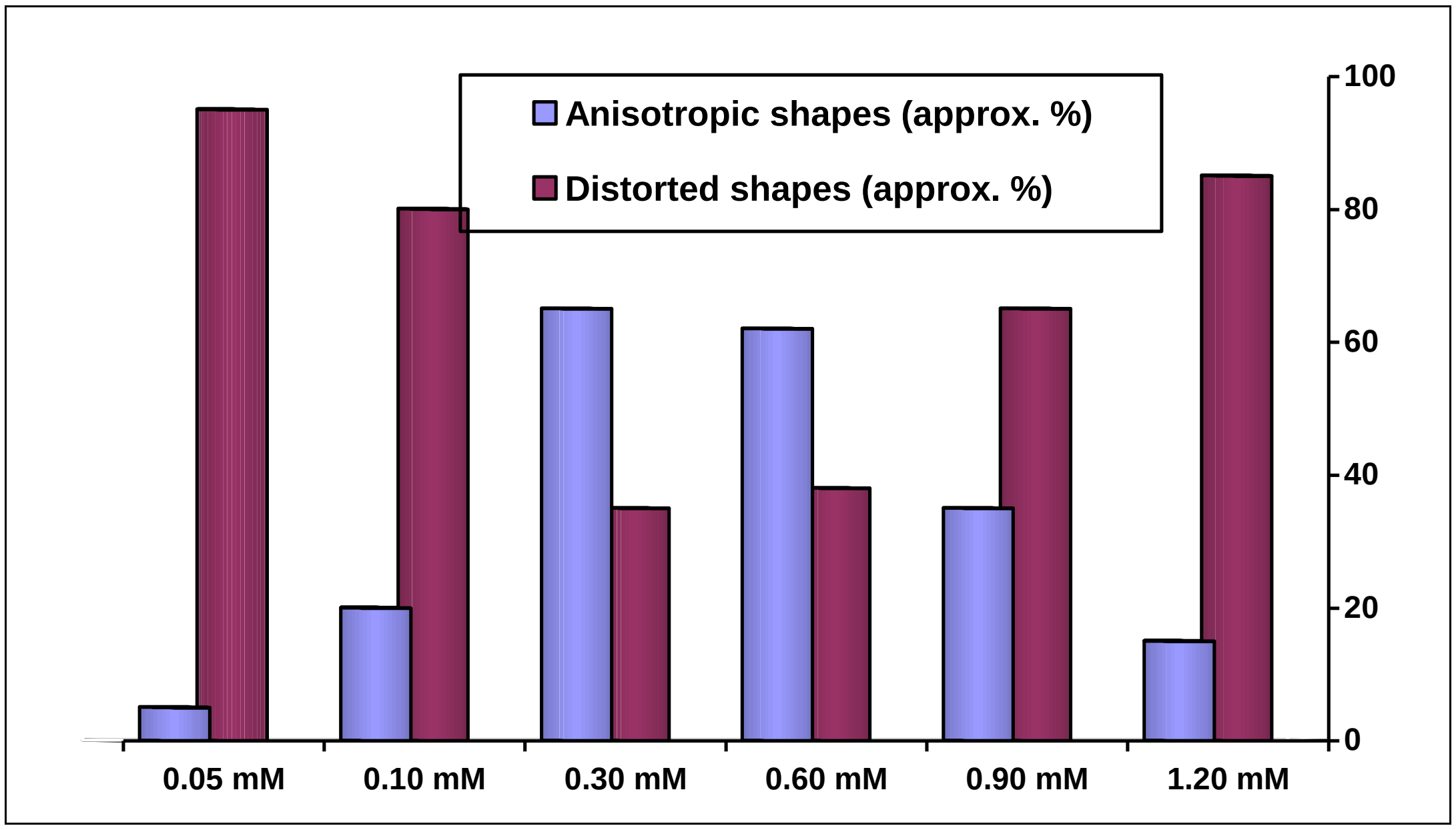

**Figure 10.** Approx. %age of anisotropic and distorted shapes developed at different concentrations of gold precursor while employing pulse-based electron-photon/solution interface process when fixed duration in each process was 10 minutes and bipolar pulse ON/OFF time was 10 µsec.

The mechanism behind the particular morphology in various anisotropic nanoparticles and particles remained the same with different precursor concentration. The difference in size is observed only. The assembling orientations of structures of smooth elements (of different-sized and **triangle**-shaped tiny particles) remained intact in the course of developing particular morphology anisotropic nanoparticles and particles. The attained orientations of packing of tiny-shaped particles for each face of these nanoparticles and particles, which remained intact during their development, can be drawn by taking their centers as the reference points. Under normal conditions of processing molar concentrations in pulse-based electron-photon/solution interface process, these anisotropic nanoparticles and particles do not involve largely any sort of interaction to alter the developed features of their morphology due to synchronization of the process. However, distorted particles may introduce this factor.

Origin of atoms to be in gas and solid states along with their different transition states has been discussed elsewhere [39]. As per the nature of atoms of tiny-sized particles, they can be a defective nanomedicine instead of being an effective one [40]. Gold shapes of one-dimension and multi-dimension clearly identify the role of forces at ground surface (in surface format) [41]. Developing hard coating is nearly under the switched force and energy behaviors of gas and solid atoms [42]. It has been pointed out that upto certain numbers of atoms, tiny particles developed in hcp structures [43] and tiny particle size upto a point shows metallic character [44]. It has been stated that research efforts that are in progress should also consider the dynamics to explain geometry and entropy in addition to structure [45]. Disordered jammed configuration is not the only one in any known protocol, but there are also ordered metrics, which characterize the order of packing [46]. A study on size-controlled gold nanoparticles synthesized in photochemical process is discussed elsewhere [47]. A recent study shows the exertion of force in **amalgamation** of nanoparticles **and** particles inside the solution [48].

From the application point of view, nanoparticles and particles having distorted shapes show potential in various catalytic applications, whereas those in anisotropic shapes indicate a potential to be used in ultra-high speed devices along with applications in diversified areas such as optics, medical and photonic devices, etc. Therefore, anisotropic nanoparticles and particles of different features enroll their strong applications for the areas, where highly controlled field is the requirement at significant diligent points. However, nanoparticles and particles of different distorted shapes reveal their strong applications in various catalytic activities. Gold products in certain sizes and shapes can be driven for different biological applications, and they can also have an impactful societal role.

## 4. Conclusion

The development of tiny particles in different sizes depends on the initial amount of precursor concentration. A certain amount of precursor concentration under the fixed ratio of bipolar pulse OFF to ON time results in the development of many tiny particles having a triangular shape. Increasing the molar concentration of gold precursor from 0.05 mM to 1.20 mM, average size of tiny particles also increases from 1.3 nm (approx.) to 50 nm (approx.). At 0.05 mM, tiny-sized particles do not develop in a triangular shape. Packing of such tiny particles results in the development of less distorted sphere-shaped nanoparticles. At 0.07 to 0.90 mM, many tiny-sized particles develop in a triangular shape, but they develop in maximum amount (number) at precursor concentration 0.30 mM and 0.60 mM.

In custom-built pulse-based electron-photon/solution interface process, at suitable precursor concentrations, the development of anisotropic gold nanoparticles and particles is due to the packing of triangle-shaped tiny particles, where made structures of smooth elements assembled at a common point under controlled orientations of electrons available at tips. Development of distorted particles is under the coalescences of non-triangular **shape** tiny particles, where they do not necessarily coalesce at a common point. At 1.20 mM, many tiny particles develop in very large size but not in triangular shape, so, as a result of their coalescences, distorted particles also develop. The SAPR patterns of distorted particles show their structure in disorderliness.

At 0.05 mM precursor concentration, nanoparticles of very small size develop that are termed as distorted or spherical **shape** nanoparticles. At 1.20 mM precursor concentration, mainly distorted particles having very large size develop. Between 0.05 mM and 1.20 M precursor concentrations, distorted shapes of particles also develop with the increase in size. Up to 0.20 mM precursor concentration, developed one- and multi-dimensional shapes of gold are mainly related to nanoparticles. For 0.30 mM precursor concentration, developed shapes of gold have size range both in nanoparticle and particle. For precursor concentration greater than 0.30 mM, developed shapes have mainly their size in the range of particle size. A tiny particle or tiny-sized particle is a first-hand resulted colloidal entity, whereas, a nanoparticle or particle is a second place resulted colloidal entity. A nanoparticle of any anisotropic shape is smaller in size than a particle, which can have length of its side around 100 nm. A tiny particle can be bigger in size than a nanoparticle where it develops at higher precursor concentration. However, it does not develop at second place. So, to differentiate it from small-sized tiny particle or nanoparticle, it can be termed as tiny particle of large size. To develop a nanoparticle, coalescence of minimum two tiny particles is required.

The decreasing Argon gas flow rate from 100 sccm to 50 sccm does not alter the morphology of nanoparticles and particles. However, change in Argon gas flow rate influences the electronic structure of atoms forming structures of smooth elements within sub-angstrom. Generally, at 50 sccm Argon gas flow rate, atoms configuring their structures of smooth elements (to develop nanoparticles and particles of anisotropic shapes) elongated less as compared to those developed at 100 sccm.

The color of solution for each processed molar concentration of gold changes due to overall impact of the incident light. The modified inter-state electron gap of elongated atoms forming structures of smooth elements in nanoparticles and particles of different shapes is the main cause of distinctive color of light.

## Acknowledgements

Mubarak Ali thanks Ministry of Science and Technology, Taiwan for awarding postdoctorship: NSC-102-2811-M-032-008 (August 2013- July 2014). Authors wish to thank Dr. Kamatchi Jothiramalingam Sankaran, National Tsing Hua University and Mr. Vic Chen, Tamkang University, Taiwan (R.O.C.) for assisting in TEM operation.

## References

[1] Daniel M–C and Astruc D 2004 *Chem. Rev.* **104** 293

[2] Brust M, Walker M, Bethell D, Schiffrin D J and Whyman R 1994 *J. Chem. Soc., Chem. Commun.* **801**.

[3] Whetten R L *et al* 1996 *Adv. Mater.* **8** 428

[4] Link, S and El-Sayed, M A 2000 *Inter. Rev. Phys. Chem.* **19** 409

[5] Brown L O and Hutchison J E 2001 *J. Phys. Chem. B* **105** 8911

[6] Whitesides G M and Boncheva M 2002 *Proc. Natl. Acad. Sci. U.S.A.* **99** 4769

[7] Brust M and Kiely C J 2002 *Colloids and Surfaces A: Physicochem. Eng. Aspects* **202** 175

[8] Huang J, Kim F, Tao AR, Connor S, Yang P 2005 *Nat. Mater.* **4** 896

[9] Glotzer S C, Horsch M A, Iacovella C R, Zhang Z, Chan E R and Zhang X 2005 *Curr. Opin. Colloid Interface Sci.* **10** 287

[10] Glotzer S C and Solomon M J 2007 *Nature Mater.* **6** 557

[11] Shaw C P, Fernig D G and Lévy R 2011 *J. Mater. Chem.* **21** 12181

[12] Vanmaekelbergh D 2011 *Nano Today* **6** 419

[13] Lofton C and Sigmund W 2005 *Adv. Funct. Mater.* **15** 1197

[14] Millstone J E, Hurst S J, Métraux G S, Cutler J I and Mirkin C A 2009 *Small* **5** 646

[15] Cuffari B 2019 https://www.azonano.com/article.aspx?ArticleID=5142

[16] Mariotti D, Patel J, Švrček V and Maguire P 2012 *Plasma Process. Polym.* **9**, 1074

[17] Patel J, Něemcová L, Maguire P, Graham W G and Mariotti D 2013 *Nanotechnology* **24** 245604

[18] Huang X, Li Y and Zhong X 2014 *Nanoscale Research Lett.* **9** 572

[19] Saito N, Hieda J and Takai O (2009) *Thin Solid Films* **518** 912

[20] Furuya K, Hirowatari Y, Ishioka T and Harata A 2007 *Chem. Lett.* **36** 1088

[21] Hieda J, Saito N and Takai O 2008 *J. Vac. Sci. Technol. A* **26**, 854

[22] Shirai N, Uchida S and Tochikubo F 2014 *Jpn. J. Appl. Phys.* **53** 046202

[23] Baba K, Kaneko T and Hatakeyama R 2009 *Appl. Phys. Exp.* **2** 035006

[24] Liu Y and Zhang X 2011 *Chem. Soc. Rev.* **40** 2494

[25] Kuzyk A *et al* 2012 *Nature* **483** 311

[26] Kim J, Lee Y and Sun S 2010 *J. Am. Chem. Soc.* **132** 4996

[27] Kusada K *et al* 2013 *J. Am. Chem. Soc.***135** 5493

[28] Ali M and Lin I –N 2019 *arXiv* http://arxiv.org/abs/1604.07144**v15**

[29] Ali M 2019 *arXiv* http://arxiv.org/abs/1611.05392**v22**

[30] Ali M 2017 *Preprints* http://www.preprints.org/manuscript/201701.0028/v10

[31] Ali M 2019 *arXiv* http://arxiv.org/abs/1611.01255v20

[32] Ali M 2019 *arXiv* http://arxiv.org/abs/1609.08047**v22**

[33] Ali M and Lin I –N 2019 *Surf. Interface Anal.* **51** 389

[34] Ali M and Ürgen M 2017 *Carbon* **122** 653

[35] Ali M, Lin I –N and Yeh C –Y 2018 *NANO* **13** 1850073

[36] Ali M and Lin I –N 2019 *Adv. Nat. Sci.: Nanosci. Nanotechnol.* **10** 025015

[37] Ali M and Lin I –N 2019 *Appl Nanosci* **9** 1367

[38] Ali M, Lin I –N and Yeh C –J 2018 *NANO* **13** 1850109

[39] Ali M 2019 https://www.researchgate.net/publication/323723379

[40] Ali M 2018 *J. Nanomed. Res.* **5** 241

[41] Ali M and Lin I –N 2019 https://www.researchgate.net/publication/329066950

[42] Ali M, Hamzah E and Toff M R M 2019 *Preprints* https://www.preprints.org/manuscript/201802.0040/v10

[43] Negishi Y *et al* 2015 *J. Am. Chem. Soc.* **137** 1206

[44] Moscatelli A 2015 *Nature Nanotechnol.* https://doi.org/10.1038/nnano.2015.16

[45] Manoharan V N 2015 *Science* **349** 1253751

[46] Atkinson S, Stillinger, F H and Torquato S 2015 *Proc. Natl. Acad. Sci. U.S.A.* **111** 18436

[47] Kim, J -H, Lavin, B W, Burnett, R D and Boote, B W 2011 *Nanotechnology* **22** 285602

[48] Ali M and Lin I -N 2019 *ChemRxiv* https://doi.org/10.26434/chemrxiv.9823625.**v2**

# Supplementary Materials

In Figure S1 (A-F), each SAPR pattern indicates the features of its associated particle. Intensity spots in each pattern show high precision in terms of order**, so** in the anisotropy of particles **also**. In the SAPR pattern when labeled distance is ~ 0.24 nm, it forms a multi-dimensional (MD) shape of the particle. In the SAPR pattern when labeled distance is ~ 0.27 nm, it forms a one-dimensional (1D) shape of the particle.

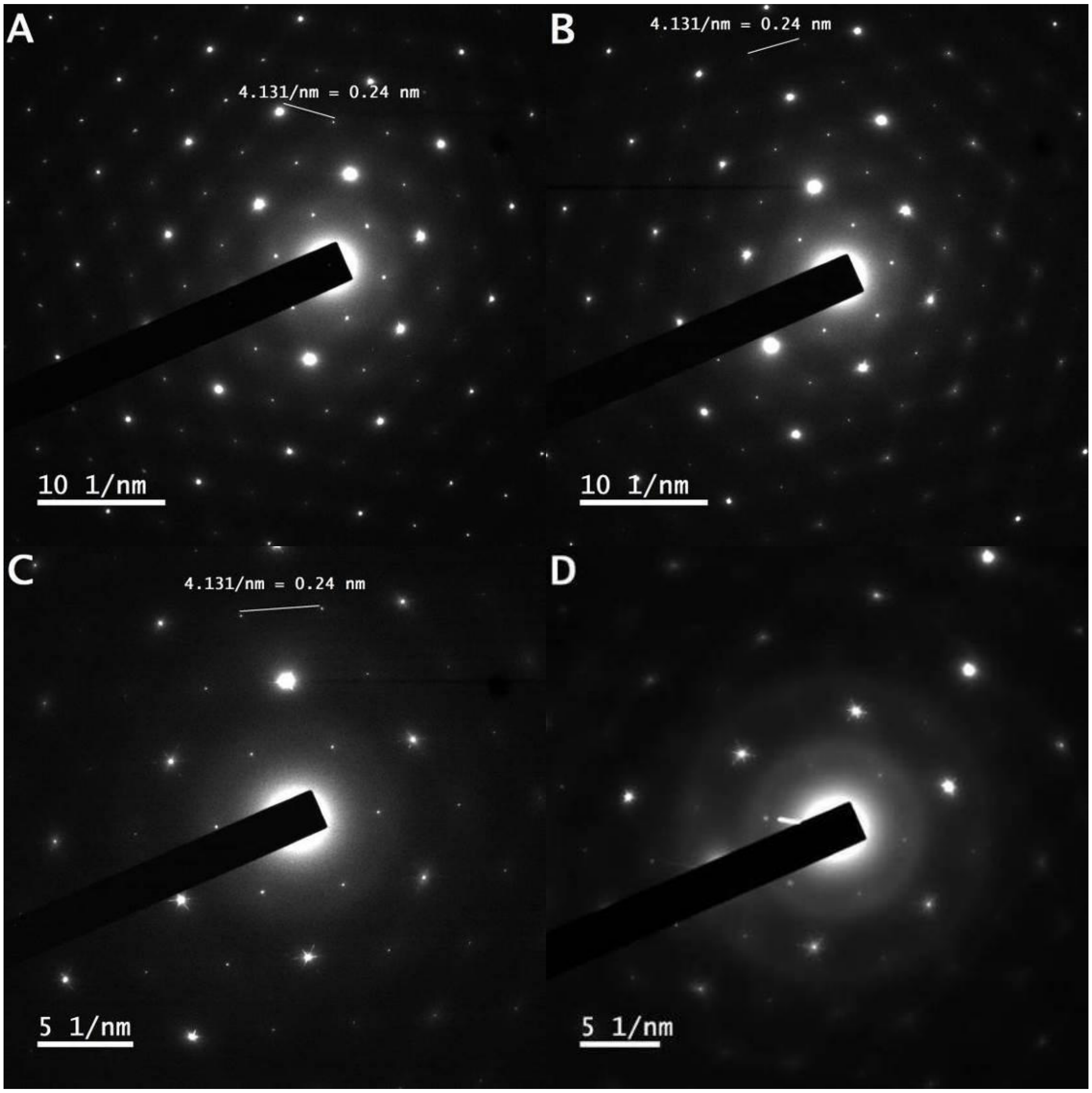

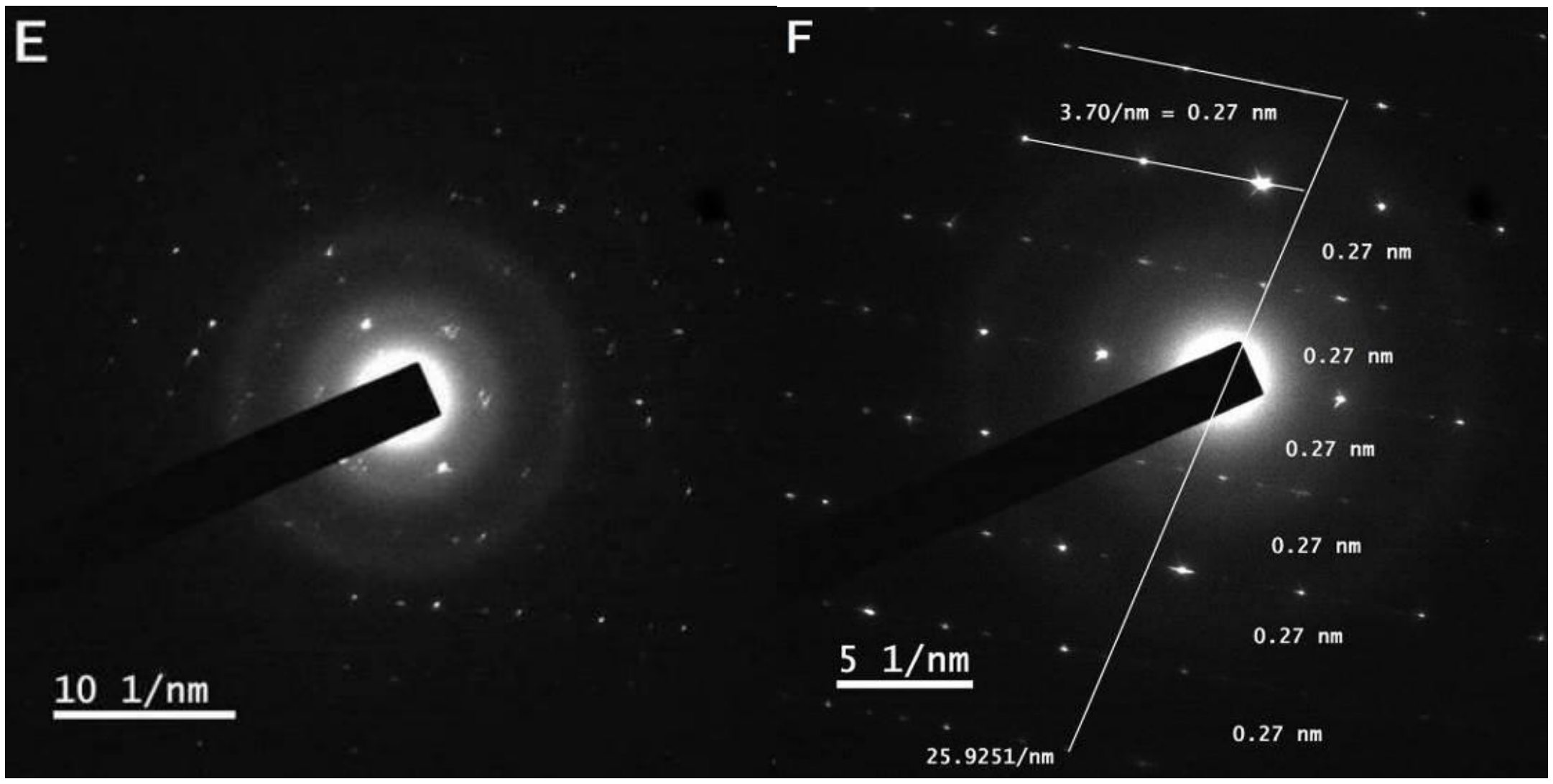

**Figure S1.** (A-F) SAPR patterns of particles shown in Figure 5 (a-f).

SAPR patterns in Figure S2 (A-F) reveal the structures of particles (shown in Figure 6a-f) in regular order to the large extent. The two different labelled distances, in the case of particles having MD shapes ~0.24 nm and in the case of particle having 1D shape ~0.27 nm, are discussed in a separate study [41]. In Figure S2 (C), SAPR pattern indicates that spotted spots reflected at the front surfaces are at a certain pace, where photons (and they are not electrons) reflected from the structures of smooth elements of above positioned particle (hexagon-shaped particle) as well as the underneath one (hexagon-shaped particle). In the latter case, photons reflected at the surface of underlying structure while entering through the inter-spacing distance of structures of smooth elements of above positioned particle. Each structure of smooth element in both **hexagon**-shaped particles was related to the elongated atoms of an array belonging to a tiny-shaped particle.

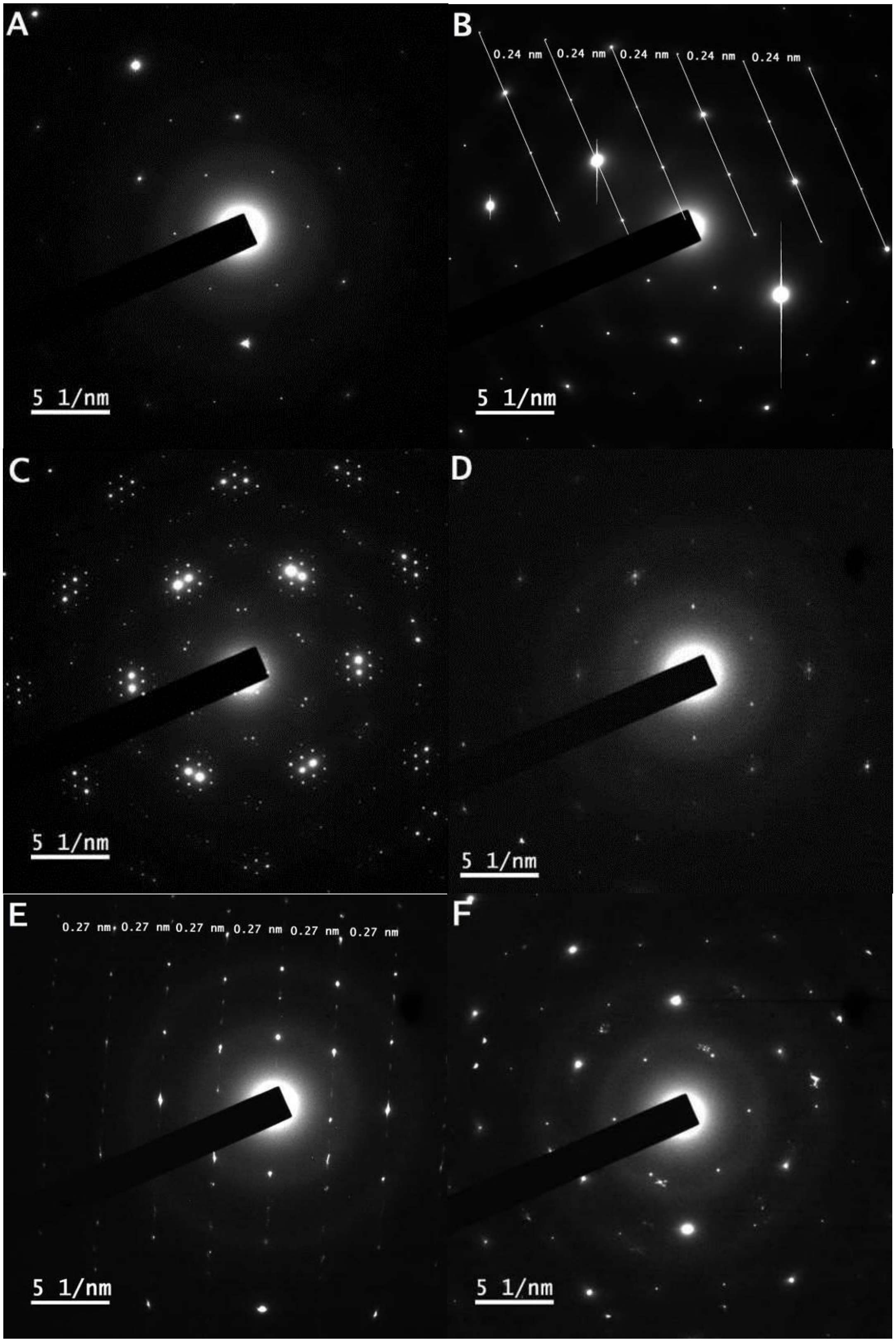

**Figure S2.** (A-F) SAPR patterns of particles shown in Figure 6 (a-f).

SAPR patterns shown in Figure S3 (A & B) show partially regular (ordered) and partially irregular (disordered) structure of their particles. SAPR patterns of distorted particles shown in Figure S3 (C & D) show highly irregular (disordered) structure. The spots of photons in the patterns reflected at the surface of particles (shown in Figures 7a & 7b) covering a few elongated atoms indicate uniform structural features (shown in Figure S3A and Figure S3B). This is because of the selected area of each pattern in only a few square nanometers. In Figure S3 (C) and Figure S3 (D), the structural features of their distorted particles are shown in Figures 7 (c) and 7 (d), distribution of spotted intensity of reflected photons at selected area is not in the repeating order.

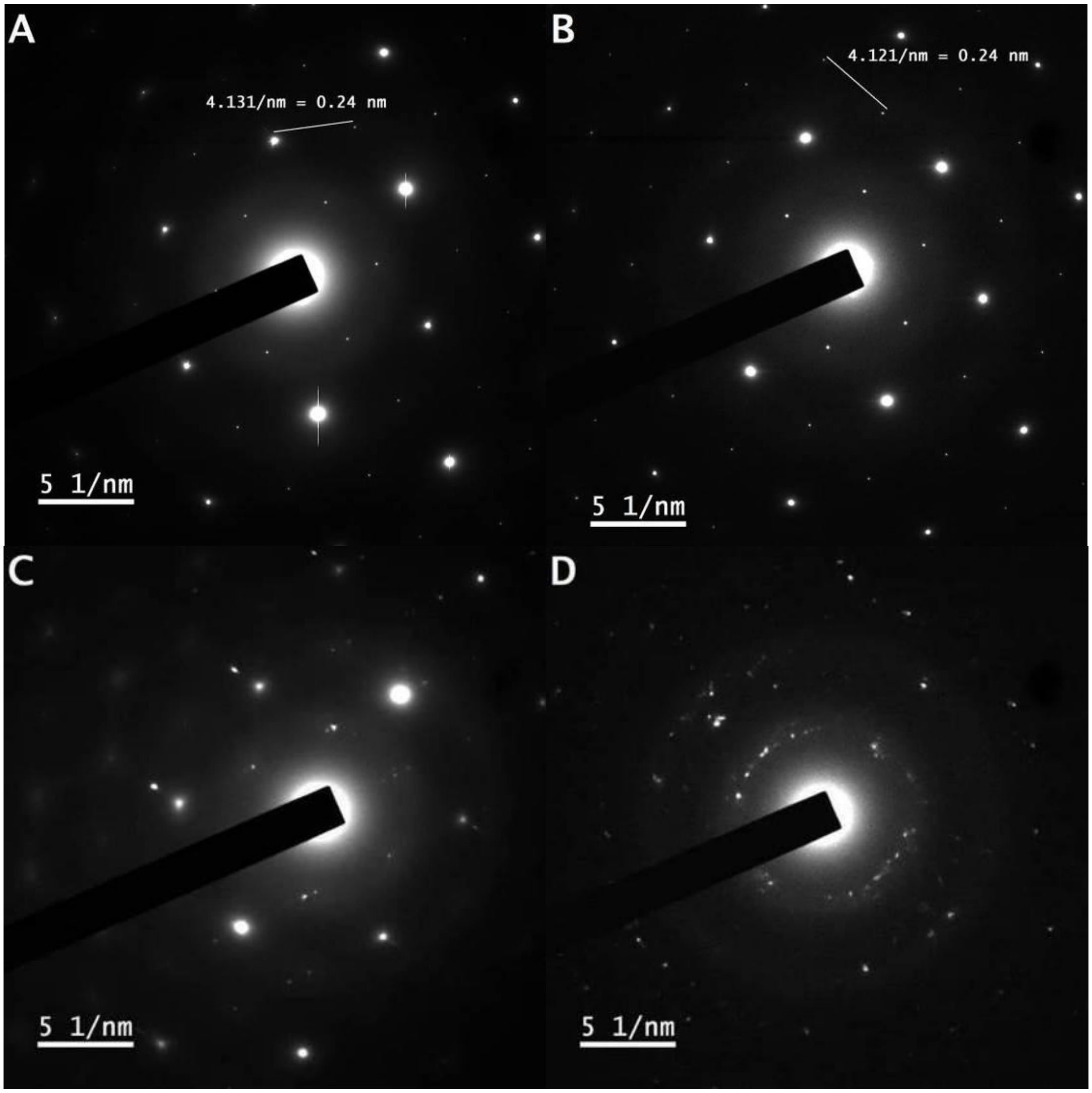

**Figure S3.** (A-D) SAPR patterns of particles shown in Figure 7 (a-d).

In SAPR patterns of particles, shapes other than rod or bar having distance between parallel printed intensity spots of ~0.24 nm are labeled in Figures S1 (A-C), Figure S2 (B) and Figure S3 (A & B). On the other hand, the distance between parallelly printed intensity spots (which are now intensity lines) in case of rod (or bar)-shaped particles is ~0.27 nm as shown in Figure S1 (F) and in Figure S2 (E). A separate study has discussed the cause of printing intensity spots of reflected photons in the shape of lines for particles of 1D shapes and in the shape of dots for particles of MD shapes [41].

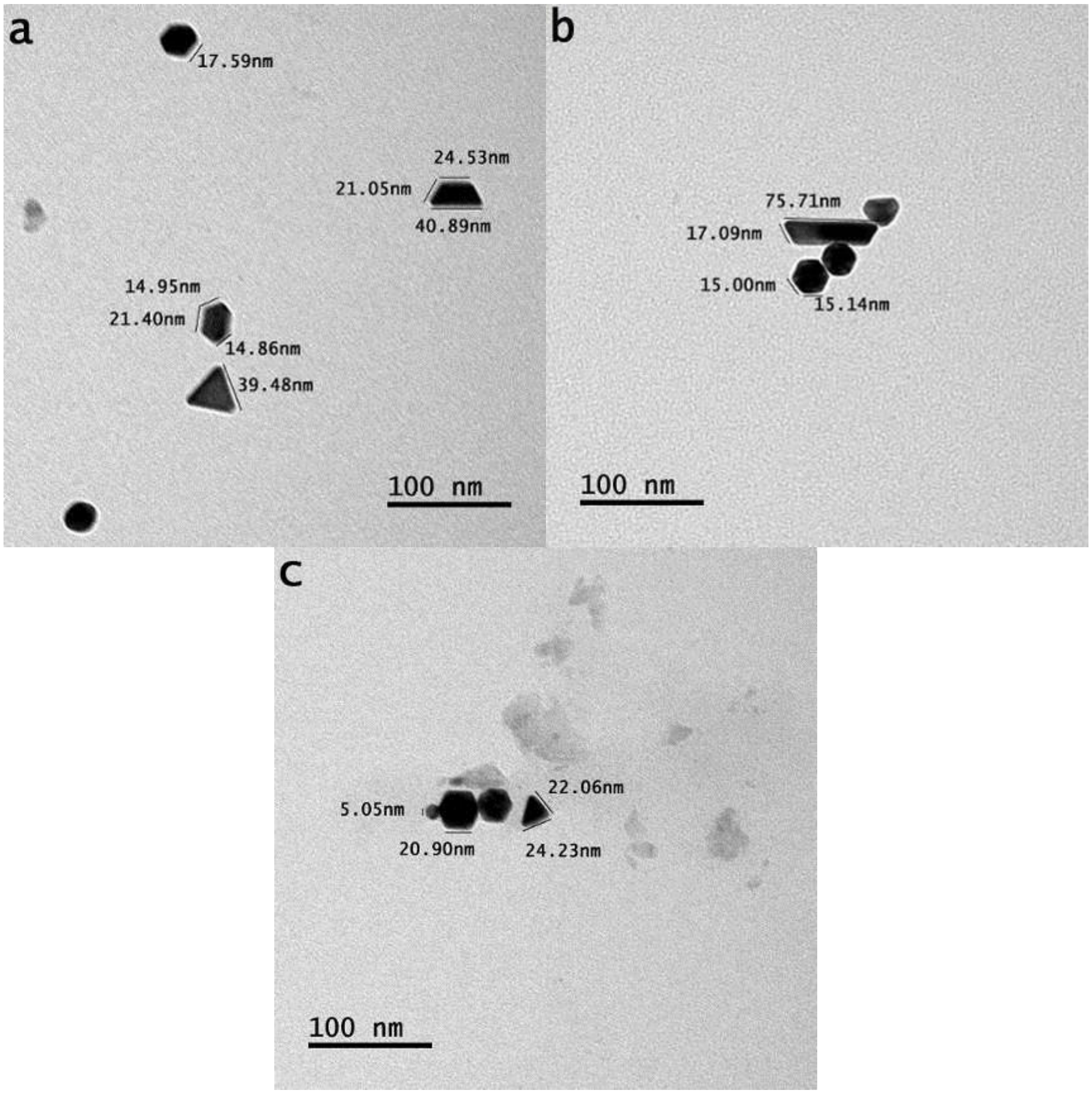

**Figure S4.** (a-c) BF-TM images of nanoparticles showing both geometrical and distorted shapes; precursor concentration 0.07 mM and Argon flow rate 50 sccm.

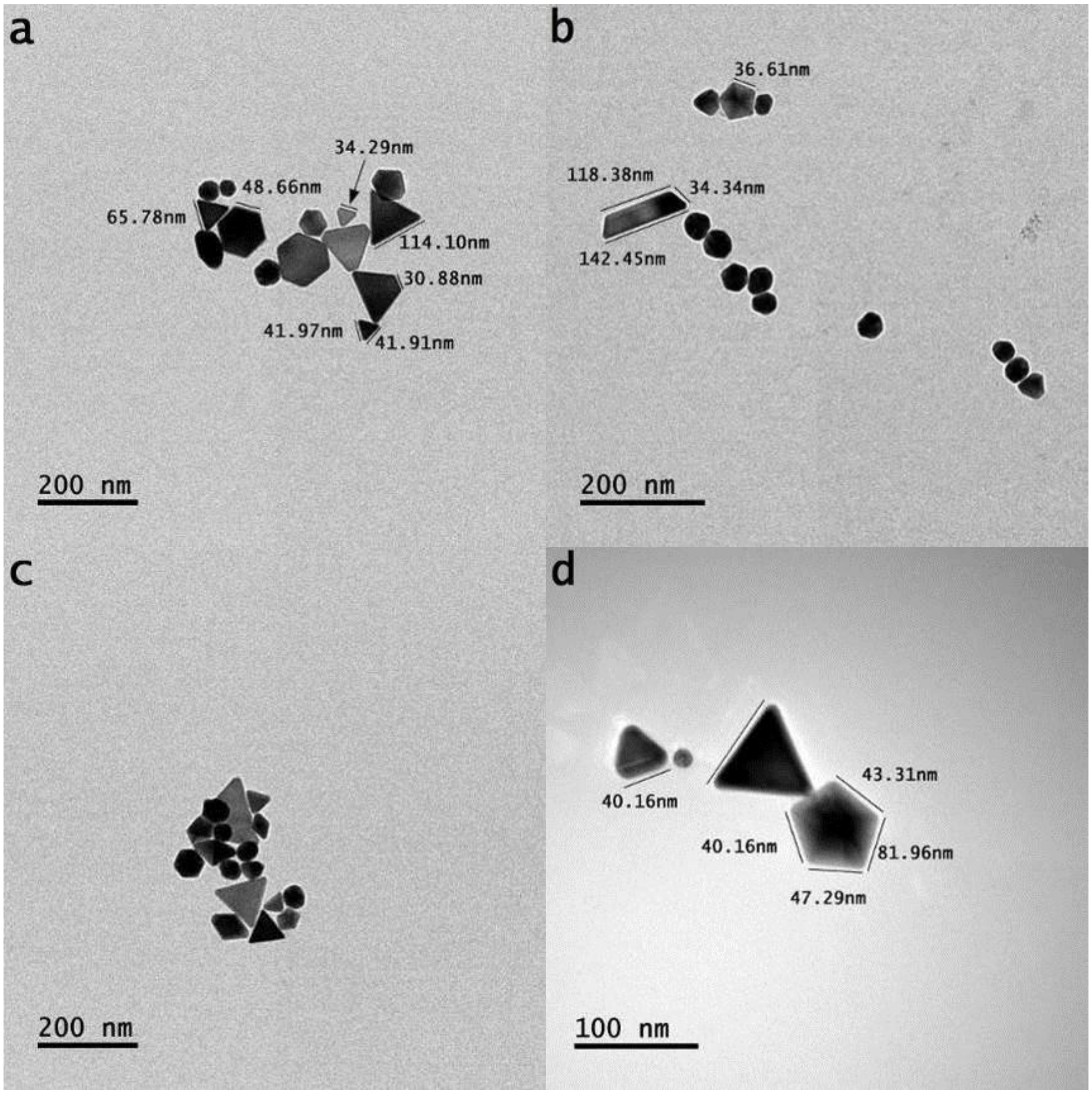

**Figure S5.** (a-d) BF-TM images of nanoparticles showing both geometrical and distorted shapes; precursor concentration 0.10 mM and Argon flow rate 50 sccm.

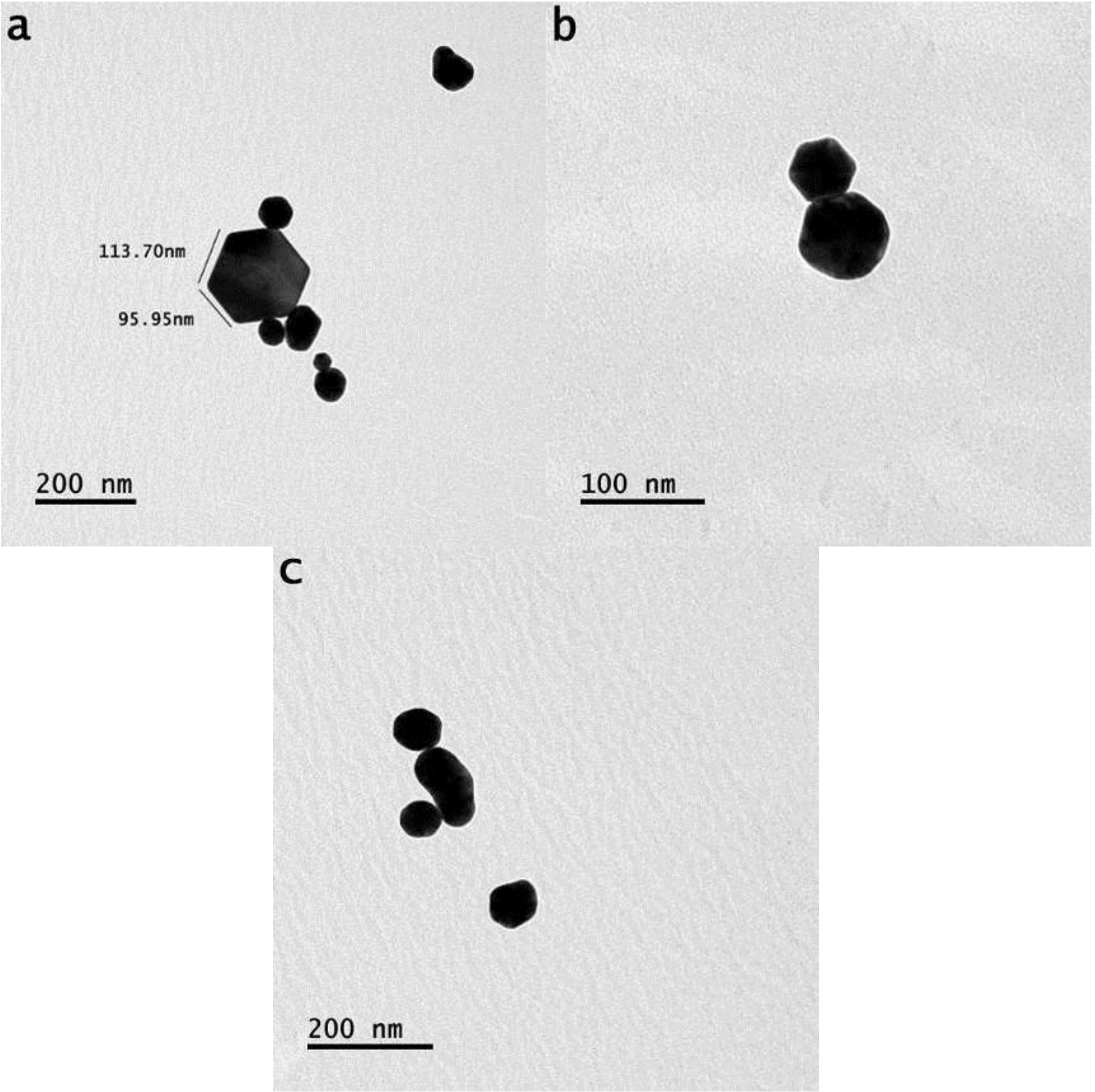

**Figure S6.** (a-c) BF-TM images of nanoparticles and particles showing both geometrical and distorted shapes; precursor concentration 0.30 mM and Argon flow rate 50 sccm.

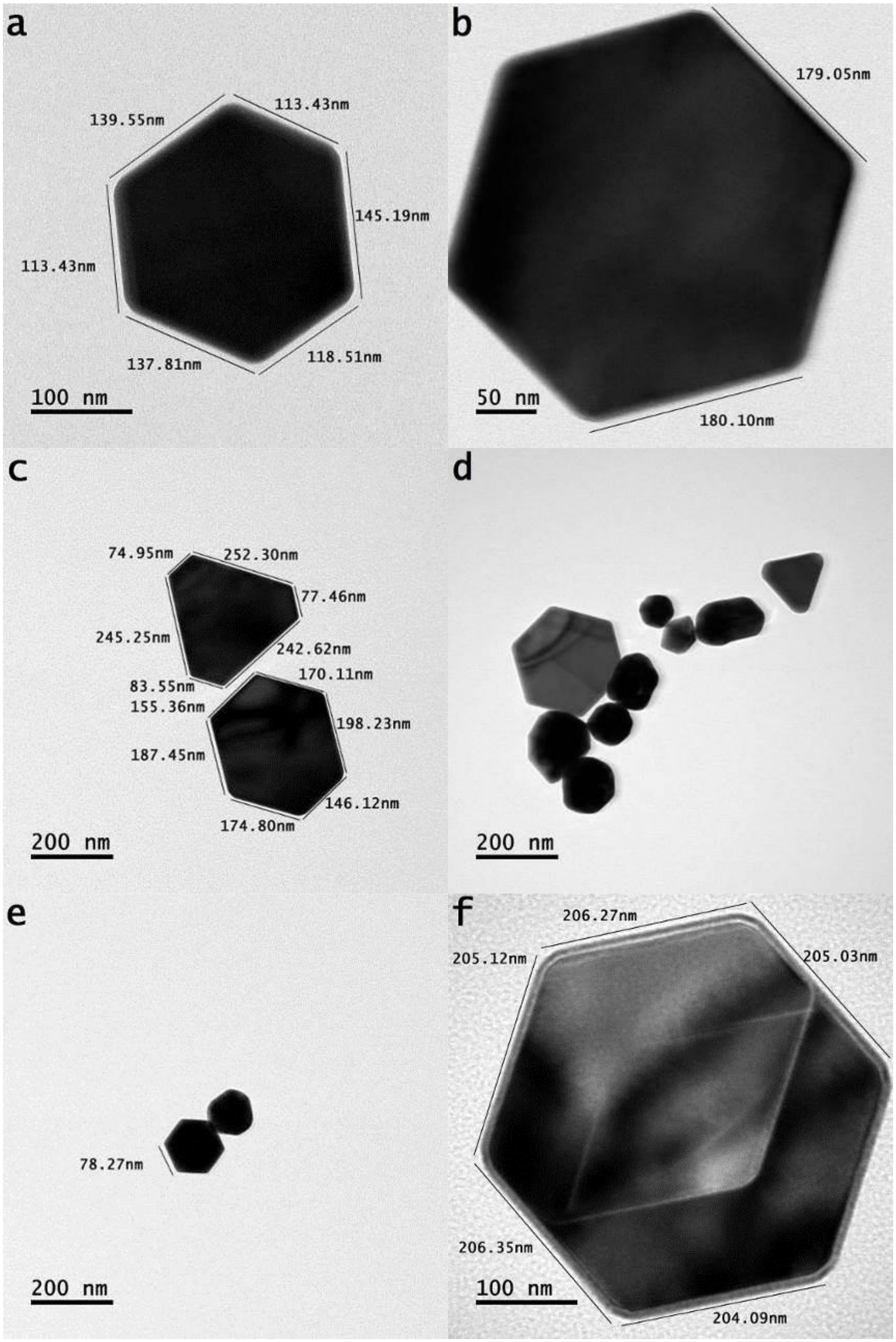

a
139.55nm
113.43nm
145.19nm
113.43nm
137.81nm
118.51nm
100 nm
b
179.05nm
50 nm
180.10nm
c
74.95nm
252.30nm
77.46nm
245.25nm
242.62nm
170.11nm
83.55nm
155.36nm
198.23nm
187.45nm
146.12nm
174.80nm
200 nm
d
200 nm
e
78.27nm
200 nm
f
206.27nm
205.12nm
205.03nm
206.35nm
100 nm
204.09nm

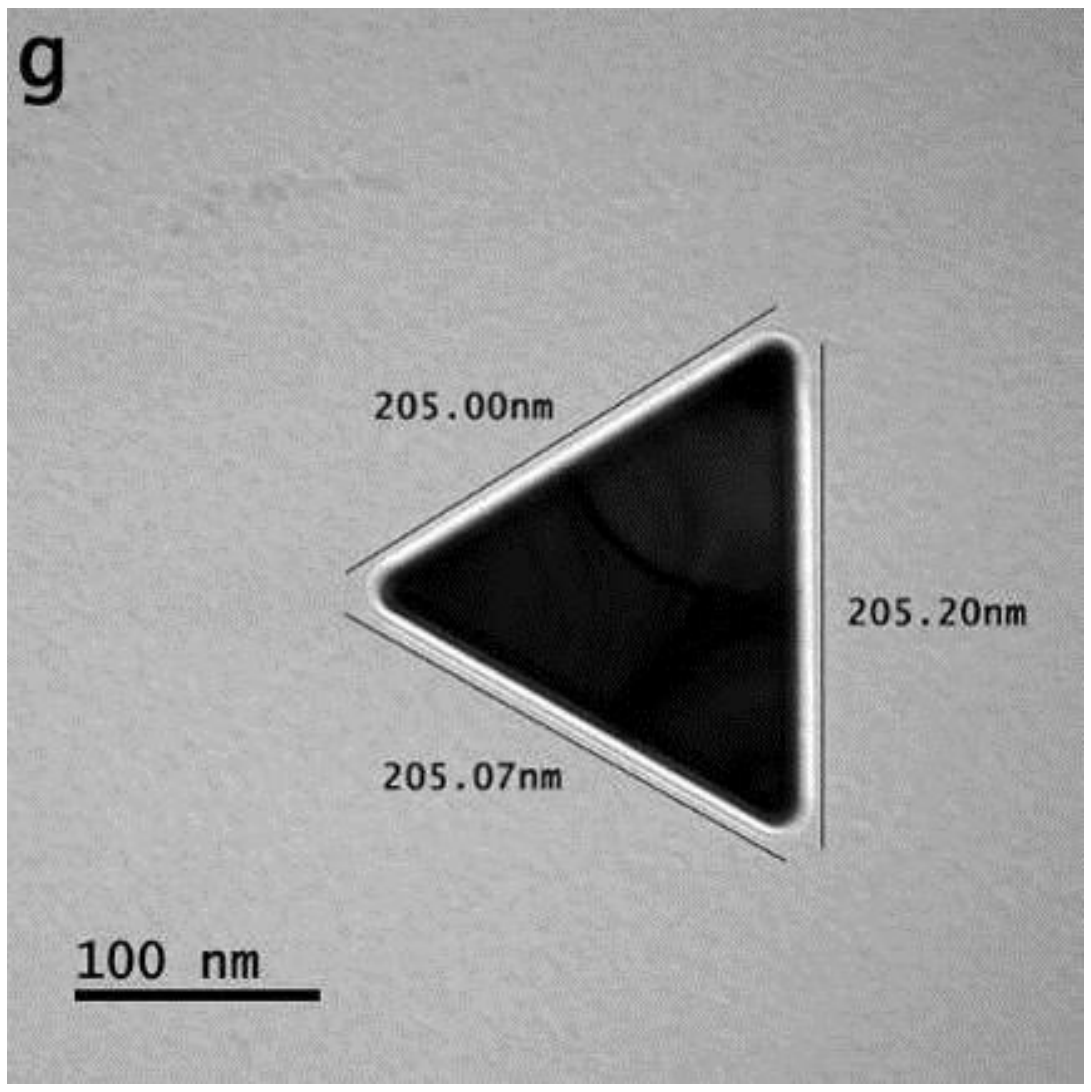

**Figure S7.** (a-g) BF-TM images of particles showing both geometrical and distorted shapes; precursor concentration 0.60 mM and Argon gas flow rate 50 sccm.

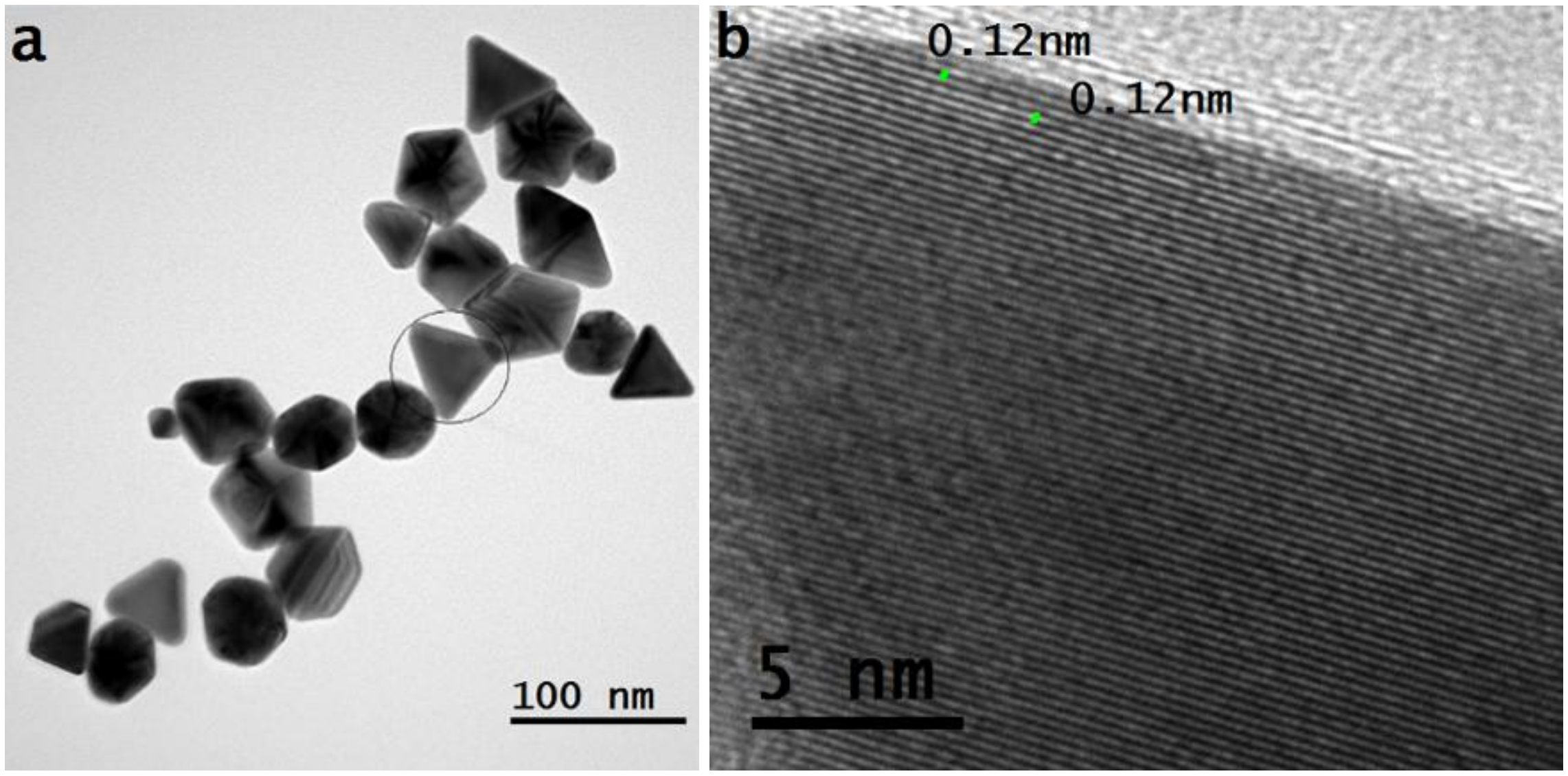

**Figure S8.** (a) BF-TM image of nanoparticles showing both geometrical and distorted shapes and (b) magnified HR-TM image taken from the one vertex of encircled triangle in 'a' shows nearly equal width of each structure of smooth element and inter-spacing distance; precursor concentration 0.10 mM and Argon gas flow rate 50 sccm.

## Authors' biography:

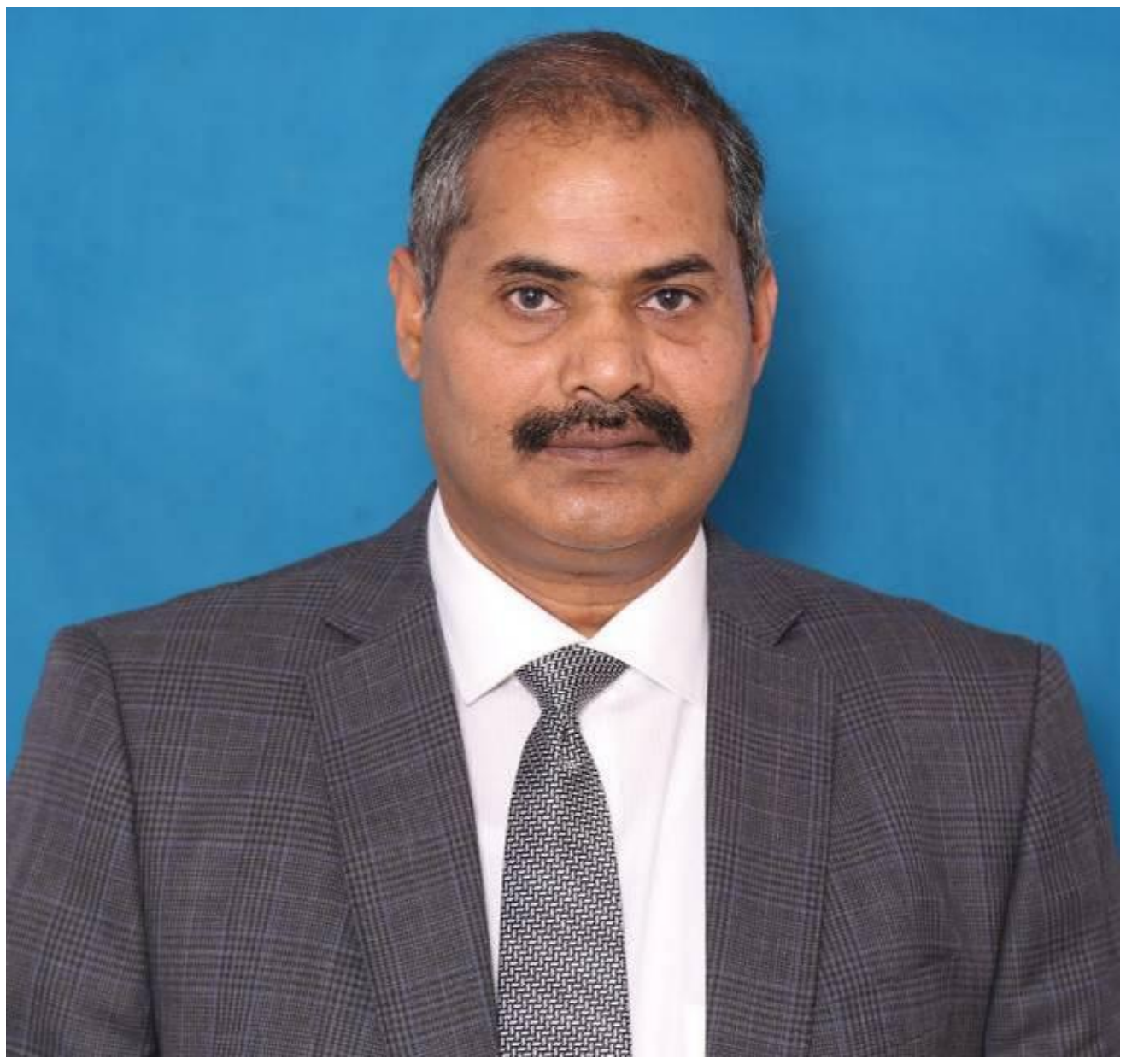

**Mubarak Ali** graduated from University of the Punjab with BSc (Phys & Maths) in 1996 and MSc Materials Science with distinction from Bahauddin Zakariya University, Multan, Pakistan (1998); his thesis work completed at Quaid-i-Azam University Islamabad. He gained PhD in Mechanical Engineering from the Universiti Teknologi Malaysia under the award of Malaysian Technical Cooperation Programme (MTCP;2004-07) and postdoc in advanced surface technologies at Istanbul Technical University under the foreign fellowship of The Scientific and Technological Research Council of Turkey (TÜBİTAK, 2010). He completed another postdoc in the field of nanotechnology at the Tamkang University Taipei (2013-2014) sponsored by National Science Council now M/o Science and Technology, Taiwan (R.O.C.). Presently, he is working as Assistant Professor on tenure track at COMSATS University Islamabad (previously known as COMSATS Institute of Information Technology), Islamabad, Pakistan (since May 2008) and prior to that worked as assistant director/deputy director at M/o Science & Technology (Pakistan Council of Renewable Energy Technologies, Islamabad, 2000-2008). He was invited by Institute for Materials Research, Tohoku University, Japan to deliver scientific talk. He gave several scientific talks in various countries. His core area of research includes materials science, physics & nanotechnology. He was also offered the merit scholarship for the PhD study by the Government of Pakistan, but he did not avail himself of the opportunity. He also earned Diploma (in English language) and Certificate (in Japanese language) in 2000 and 2001 respectively, in part-time from the National University of Modern Languages, Islamabad. He is the author of several articles available at following links; https://scholar.google.com.pk/citations?hl=en&user=UYjvhDwAAAAJ, https://www.researchgate.net/profile/Mubarak_Ali5, https://www.mendeley.com/profiles/mubarak-ali7/, & https://publons.com/researcher/2885742/mubarak-ali/publications/

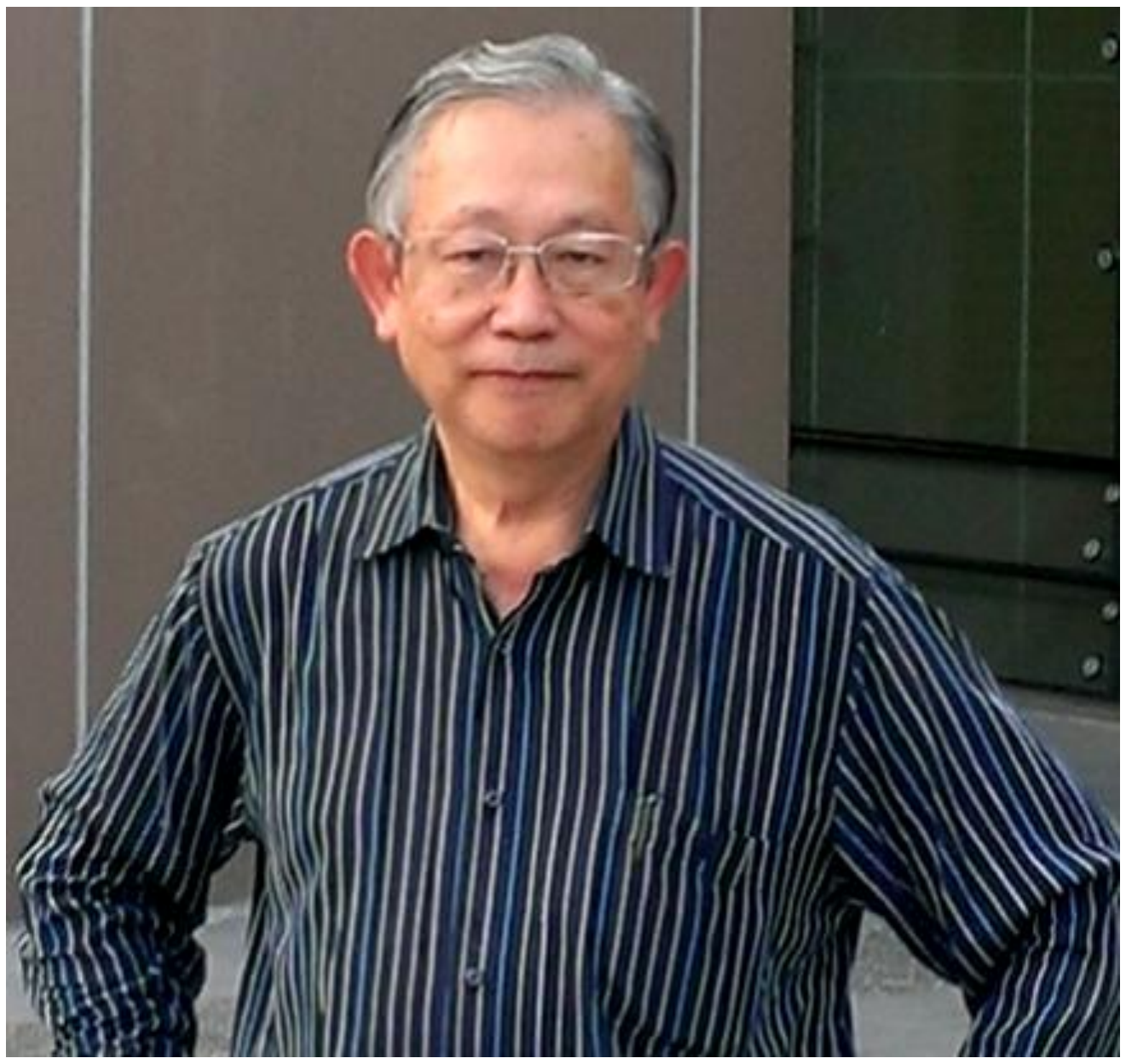

His Excellency, **I-Nan Lin**, worked as a **Senior Professor** for several years at Tamkang University, Taiwan, and now **Professor Emeritus**. His Excellency received the bachelor's degree in physics from National Taiwan Normal University, Taiwan, M.S. from National Tsing-Hua University, Taiwan, and the Ph.D. degree in Materials Science from U. C. Berkeley in 1979, U.S.A. His Excellency, worked as a senior researcher in Materials Science Centre in Tsing-Hua University for several years and a faculty member as senior professor in Department of Physics, Tamkang University. Professor Lin published several hundred referred journal publications and held distinctive position in university. Professor I-Nan Lin supervised several PhD and Postdoc candidates around the world. Professor I-Nan Lin remained researching in the areas of developing high conductivity diamond films, composite materials and the deep-rooted studies of the transmission microscopy of materials.